\documentclass{article}

\usepackage[utf8]{inputenc}     
\usepackage[margin=10pt,font=small,labelfont=bf,labelsep=endash]{caption}
\usepackage[T1]{fontenc}
\usepackage{setspace}
\usepackage{booktabs}
\usepackage{multirow}
\usepackage{tabularx}
\usepackage{paralist}
\usepackage{ifthen}
\usepackage{listings}
\usepackage{amsmath}
\usepackage{amsfonts}
\usepackage{color}
\usepackage[table]{xcolor}
\usepackage{algorithm}
\usepackage{algpseudocode}
\usepackage{url}
\usepackage{rotating}
\usepackage{subfig}
\usepackage{arydshln}
\usepackage{todonotes}
\usepackage[normalem]{ulem}
\usepackage{siunitx}
\usepackage{authblk}
\usepackage{hyphenat}
\usepackage[absolute]{textpos}

\newcommand{\algfunc}{\texttt}

\newcommand\javaclass[1]{{\normalfont\fontfamily{cmvtt}\selectfont #1}}
\newcommand\javainterface[1]{{\itshape\fontfamily{cmvtt}\selectfont #1}}

\makeatletter
\@ifpackageloaded{algorithm}
	{\algrenewcommand\algorithmicforall{\textbf{for each}}}
	{}
\@ifpackageloaded{amsmath}{
	\DeclareMathOperator*{\argmax}{arg\,max}
	\DeclareMathOperator*{\argmin}{arg\,min}}
	{}
\makeatother

\newcommand*\mean[1]{\overline{#1}}

\title{Parallelization Strategies for Spatial Agent-Based Models}

\author[1]{Nuno Fachada}
\author[2]{Vitor V. Lopes}
\author[3]{Rui C. Martins}
\author[1]{Agostinho C. Rosa}

\affil[1]{Institute for Systems and Robotics, LARSyS, Instituto Superior Técnico, Universidade de Lisboa, Lisboa, Portugal}
\affil[2]{UTEC - Universidad de Ingeniería \& Tecnología, Lima, Jr. Medrano Silva 165, Barranco, Lima, Perú}
\affil[3]{INESC TEC, Campus da FEUP, Rua Dr. Roberto Frias, 4200-465 Porto, Portugal}

\providecommand{\keywords}[1]{\textbf{\textit{Keywords---}} #1}

\date{}

\begin{document}
	
\begin{textblock*}{210mm}(3mm,3mm)
	\noindent The peer-reviewed version of this paper is published in the International Journal of Parallel Programming (\url{http://dx.doi.org/10.1007/s10766-015-0399-9}). This version is typeset by the authors and differs only in pagination and typographical detail.
\end{textblock*}

\maketitle

\begin{abstract}

Agent\hyp{}based modeling (ABM) is a bottom\hyp{}up modeling approach, where each entity of the system being modeled is uniquely represented as an independent decision\hyp{}making agent. Large scale emergent behavior in ABMs is population sensitive. As such, the number of agents in a simulation should be able to reflect the reality of the system being modeled, which can be in the order of millions or billions of individuals in certain domains. A natural solution to reach acceptable scalability in commodity multi\hyp{}core processors consists of decomposing models such that each component can be independently processed by a different thread in a concurrent manner. In this paper we present a multithreaded Java implementation of the PPHPC ABM, with two goals in mind: 1) compare the performance of this implementation with an existing NetLogo implementation; and, 2) study how different parallelization strategies impact simulation performance on a shared memory architecture. Results show that: 1) model parallelization can yield considerable performance gains; 2) distinct parallelization strategies offer specific trade\hyp{}offs in terms of performance and simulation reproducibility; and, 3) PPHPC is a valid reference model for comparing distinct implementations or parallelization strategies, from both performance and statistical accuracy perspectives.

\end{abstract}


\section{Introduction}

Agent-based modeling (ABM) is a bottom-up modeling approach, where each entity of the system being modeled is uniquely represented as an independent decision-making agent. When prompted to act, each agent analyzes its current situation (e.g. what resources are available, what other agents are in the vicinity), and acts accordingly, based on a set of rules (e.g. if-then-else rules, differential equations, neural networks, genetic algorithms, etc). These rules incorporate knowledge or theories about the respective low-level components. The global behavior of the system then emerges from the simple, self-organized local relationships between the agents \cite{NunoFachada2008}. As such, ABM is a useful tool in simulating and exploring systems that can be modeled in terms of interactions between individual agents, for example, biological cell cultures, ants foraging for food or military units in a battlefield. 

Spatial agent\hyp{}based models (SABMs) are a subset of ABMs in which a spatial topology defines how agents interact \cite{shook2013communication}. For example, an agent may be limited to interact with  agents located within a specific radius, or may only move to a near physical or geographical location \cite{macal2010tutorial}. SABMs have been extensively used to study a range of phenomena in the biological and social sciences \cite{isaac2011,shook2013communication}. 

Large scale emergent behavior in ABMs is population sensitive. As such, it is preferable that the number of agents in a simulation is able to reflect the reality of the system being modeled \cite{lysenko2008framework,keenan2012novel,husselmann2008spatial}; otherwise, the expected or desired behavior may not be observable, and model validation becomes difficult \cite{husselmann2008spatial,gulyas2009modeler}. This means that in domains such as social modeling, ecology, and biology, systems can contain millions or billions of individuals \cite{d2007sugarscape,keenan2012novel,Perumalla2010,d2009data}; consequently, simulating realistic models will involve as much agents being processed per time step \cite{d2009data}. Such large scale simulations generate a very high demand for computing power \cite{gulyas122011tools} and are impractical on typical ABM frameworks such as NetLogo \cite{wilensky1999} or Repast \cite{north2013complex}, which execute serially on the CPU \cite{d2007sugarscape,d2009data}. Additionally, stochastic models in general and ABMs in particular usually require various input parameters, which can have a range of different values. Large\hyp{}scale computational experiments are required for exploring the parameter space of such models \cite{gulyas122011tools,hill2013distribution,tang2014global}. These requirements stretch, and many times surpass, what typical off\hyp{}the\hyp{}shelf computing systems can offer, especially if models are implemented in a way to only make use of one processing element (PE), such as a CPU core. Considering that commodity processors, such as GPUs and multi\hyp{}core CPUs, are nowadays composed of several PEs, a natural solution to reach acceptable scalability in ABMs consists of decomposing models such that each component can be independently processed by a logical processor (LP\footnote{In shared memory architectures, LPs are usually represented by threads, which communicate via synchronized access to shared variables. In distributed memory scenarios, LPs are commonly represented by processes, which communicate via message passing.}) in a concurrent manner \cite{gulyas2009modeler,tang2011parallelgpu,voss2010scalable,shook2013communication,tang2009hpabm,Collier2011,tang2014global}. There are, however, two main issues when parallelizing ABMs.

The first major issue concerns communication between model components and the bottleneck it creates, which is a major limiting factor in scaling parallel SABMs \cite{tang2009hpabm,shook2013communication}. This is especially true in distributed memory scenarios, where different computational cores may be located in separate, often geographically distant, nodes \cite{gulyas2009modeler}. Communication costs may suppress the potential gains of using multiple nodes and their associated resources \cite{voss2010scalable}. Many strategies and methods have been developed to manage and reduce communication in distributed memory SABMs \cite{scheutz2006adaptive,gulyas2009modeler}, nonetheless this is still a topic of active research \cite{shook2013communication}. Whatever the scenario, model partitioning should guarantee that each model component is as independent as possible in order to minimize communication between LPs. Furthermore, communication strategies should be designed to avoid deadlocks and to preserve the causality of simulation events while efficiently exploiting parallelism \cite{hill2013distribution,shook2013communication}. 

The second major issue when parallelizing ABMs is that it is very easy to inadvertently introduce changes which modify the model dynamic. This is akin to model replication, which is not a straightforward process \cite{edmonds2003replication,wilensky2007making}. ABMs are very sensitive to implementation details: the impact that seemingly unimportant aspects such as data structures, algorithms, discrete time representation, floating point arithmetic or order of events can have on results is tremendous \cite{wilensky2007making,merlone2008}. The situation becomes more difficult with model parallelization, which by definition requires considerable changes in many of these aspects. Parry and Bithell \cite{parry2012large} provide an informative account in which they were unable to successfully replicate a serial model when converting it to a parallel one. 
Conceptual models should be well specified and adequately described in order to achieve a successful model replication. As such, some authors have suggested that there should be a minimum standard for model communication, which should include at the very least: a) a structured natural language description using formal protocols such as ODD \cite{grimm2010odd}; and, b) the model's source code, given that it is the model's definitive implementation, not subject to the vagueness and uncertainty possibly associated with verbal descriptions \cite{wilensky2007making,muller2014standardised}.

In this paper we present a multithreaded Java implementation of the PPHPC\footnote{Predator\hyp{}Prey for High\hyp{}Performance Computing} model \cite{fachada2015template}, featuring several user\hyp{}selectable parallelization schemes. The goals of this investigation are two\hyp{}fold: 1) compare the performance of the implementation presented here, realized in a ``real'' programming language, with the canonical NetLogo implementation discussed in reference \cite{fachada2015template}; and, 2) study how different parallelization strategies impact simulation performance on a shared memory architecture. Care is taken so that all the parallelization strategies of the Java implementation yield similar statistical behavior as the NetLogo version, and among themselves. Leveraging on the fact that the PPHPC model captures important characteristics of SABMs, such as agent movement and local agent interactions, several conclusions on SABM parallelization techniques and the usefulness of PPHPC as a valid reference model are drawn.

The rest of the paper is organized as follows. First, in section \ref{sec:relwork}, previous work about parallelization of SABMs in shared memory architectures is discussed. Next, section \ref{sec:methodology}, Methodology, is composed of three parts: 1) an overview of the PPHPC model; 2) a description of the Java implementation, including the several parallelization strategies; and, 3) a discussion on how to compare multiple PPHPC versions. Results, section \ref{sec:results}, show that: 1) model parallelization and the use of a ``real'' programming language (as opposed to the NetLogo modeling language) can yield considerable performance gains; 2) distinct parallelization strategies offer specific trade\hyp{}offs in terms of performance and simulation reproducibility; and, 3) PPHPC is a valid reference model for comparing distinct implementations or parallelization strategies, from both performance and statistical accuracy perspectives. Section \ref{sec:conclusions} provides a global outline of what was accomplished in this work.

\section{Background}
\label{sec:relwork}


Parry and Bithell \cite{parry2012large} describe two techniques for partitioning SABM components across multiple computational cores: agent\hyp{}parallel (AP) and environment\hyp{}parallel (EP). While these are not mutually exclusive, they are nonetheless a good starting point for reasoning about SABM partitioning. In the AP approach, the model is divided at the agent\hyp{}level, i.e. each LP is responsible for handling a set of agents. Load balancing is simpler as agents can be equally distributed among LPs so that each LP has a similar share of the computation \cite{cosenza2011distributed,parry2012large}. However, in a moving agents scenario, this partitioning leads to extra communication between LPs, which is required in order to ensure that spatially localized agent interactions are dealt with consistently, as co\hyp{}location on a LP does not guarantee co\hyp{}location in space \cite{parry2012large}. In EP partitioning, model decomposition occurs at the spatial environment level, i.e. each LP is assigned a location, together with the agents it contains \cite{cosenza2011distributed}. As such, local agent interactions will mostly occur in the same LPs. Unfortunately, when agent density varies spatially over time, e.g. in flocking or grouping patterns, load balancing issues may occur \cite{cosenza2011distributed,parry2012large,goldsby2013multithreaded}. A corner case of this issue is when simulating chemotaxis\hyp{}like patterns, where agent movement is influenced by a chemical concentration gradient, which can result in millions of agents swarming to the same location \cite{Fachada2009}.

Most attempts at parallelizing ABMs found in the literature are based on the distributed memory programming model \cite{scheutz2006adaptive,shook2013communication}, including a few generic ABM frameworks for high\hyp{}performance computing \cite{tang2009hpabm,gulyas122011tools,Collier2011,Coakley2012}. This approach allows models to scale to thousands of cores, usually found in supercomputer\hyp{}type setups \cite{yokokawa2011k,yang2011tianhe}. However, communication issues for larger models \cite{shook2013communication} and a more complex programming paradigm (when compared with multithreading on shared memory architectures) \cite{lenoski1991scalable} can restrict this approach.

Recently, the trend has been on hybrid \cite{takahashi2006efficient,aaby2010efficient}, GPU \cite{lysenko2008framework,richmond2010high,wang2012gpu,chen2015parallel} and heterogeneous \cite{vigueras2013accelerating,wang2013accelerating} methods. While these approaches allow concrete gains in simulation performance on commodity hardware, they come with an increased cost in implementation time due to the substantial more complex programming models. Hybrid methods, combining distributed and shared memory programming models, require modelers to master both paradigms, as well as specific multi\hyp{}level model decomposition. GPU architectures require the reformulation of ABMs in terms of stream SIMD\footnote{Single instruction, multiple data} computation and offer limited control flow constructs  \cite{lysenko2008framework}. Heterogeneous methods, in which both CPU and GPU are utilized, entail complex synchronization and data transfers between the two processors, also requiring careful model decomposition so that components can be efficiently processed.

Among the possible parallelization techniques, multithreading is arguably the simplest to implement \cite{lenoski1991scalable,voss2010scalable}, with the added bonus of portability. For example, modern threading APIs, such as OpenMP\footnote{\url{http://openmp.org/}} for C, C++ and Fortran or the Java 5.0 concurrency API \cite{goetz2006java}, greatly simplify multithreaded ABM implementations, and are available for a number of different shared memory CPU architectures and operating systems. In this work we focus on multithreaded SABMs implemented on traditional shared\hyp{}memory architectures.

While the majority of ABM toolkits \cite{Tobias2004,railsback2006agent,berryman2008review,nikolai2009} are targeted for single\hyp{}threaded execution on the CPU \cite{richmond2009cellular,goldsby2013multithreaded}, there have been explicit attempts to parallelize some of them. Goldsby and Pancerella \cite{goldsby2013multithreaded} describe adjustments made to the MASON agent\hyp{}based simulation package \cite{luke2005mason} that allow the use of multiple threads without major changes to conventional agent\hyp{}based programming. RepastJ \cite{north2006experiences} has adaptations to multi\hyp{}core CPUs \cite{NunoFachada2008}, while Repast Simphony \cite{north2013complex} supports parallel execution at the scheduling mechanism level. However, in most cases, the modeler must implement correct access semantics to shared data (e.g. environment and agent\hyp{}agent interaction). One of the main problems in retrofitting parallelism to existing ABM frameworks and developing new ``pure'' parallel ABMs toolkits concerns the implementation of direct agent\hyp{}to\hyp{}agent memory access, which is model dependent and requires synchronization semantics such as locks or semaphores. These constructs are provided by the threading APIs, but efficient and thread\hyp{}safe coordination of concurrent accesses still requires careful coding in order to obtain proper speedups with the number of cores, while avoiding common multithreading issues such as priority inversions or deadlocks \cite{goetz2006java}. 

EcoKit, a simulation system for spatially\hyp{}explicit ecological models \cite{Glass1997}, was one of the first parallel realizations of an ABM on a shared memory architecture, implementing a static EP solution. The authors tested the system with a mouse migration model with \num{10000} agents in a discrete 2D grid with \num{20000} cells on an eight\hyp{}processor SGI PowerChallenge machine \cite{powerchallenge1994}. One of the processors was used for the simulation kernel, while the remaining ones for the simulation itself. Results showed speedup stabilization at about four processors, reaching a maximum of \num{2.8} for seven processors.

A multithreaded SABM of immune system dynamics, parallelized using a static EP approach, is presented in reference \cite{NunoFachada2008}. Adequate speedups are obtained when chemotaxis is not simulated. However, in simulations involving this phenomena, agents in the order of thousands where shown to group in very few locations, causing severe load imbalances and limiting the scalability of simulations.

A cellular automata (CA) based ABM of opinion exchange, developed in C++ and parallelized with OpenMP, is presented by Gong et al. \cite{gong2013parallel}. The authors analyze how the performance of the parallel model varies with the size of the 2D simulation grid and with the range of agent interactions. Parallelization of the model is achieved by partitioning the CA grid in a row\hyp{}wise block\hyp{}striped fashion, with each block assigned to one thread (and consequently, to one CPU core); because the agents are fixed, and there is one agent per grid cell, this type of partitioning is both AP (agent\hyp{}parallel) and EP (environment\hyp{}parallel). Several tests were performed in a 32\hyp{}core CPU (AMD Opteron), with varying number of threads (from 1 to 32), grid sizes (up to $\num{5000} \times \num{5000}$) and interaction ranges (from \numrange{2}{1000} cells). Increasing the number of threads resulted in improved speedup, but lower efficiency (speedup divided by the number of threads). Lower efficiency also occurred when increasing the size of the simulation grid (maintaining the interaction range) and when increasing the interaction range. Nonetheless, the achieved speedups are impressive. For example, for a grid of $\num{4000} \times \num{4000}$, and an interaction range of 2 cells, a $25 \times$ speedup is obtained with 32 threads. 

Goldsby and Pancerella \cite{goldsby2013multithreaded} proposed to retrofit multithreading in the MASON ABM toolkit using an EP approach, where each thread is assigned an equal part of the simulation environment. Each simulation step is divided into two parts separated by a barrier that synchronizes access to shared environment data. Agent movement is guaranteed by interprocessor message queues. Tests were performed on a machine with 32 Intel Xeon processors using two classic ABMs: 1) Flockers, a Boids\hyp{}like model \cite{Reynolds:1987:FHS:37402.37406} in which agents display flocking behavior; and, 2) the HeatBugs model \cite{wilensky2004}. The latter was shown to scale well, particularly for larger number of agents, with effective speedup up to 28 threads. The Flockers model did not scale as well, with the authors suspecting that the tendency of agents to move in flocks lead to load balancing issues. 

The accurate characterization of the dynamics of bacterial networks led to the design of BNSim \cite{wei2013efficient}. A multithreading approach allows BNSim to efficiently simulate large populations of bacteria. As in reference \cite{goldsby2013multithreaded}, simulation steps are divided in two parts with barrier synchronization, in a tick\hyp{}tock pattern. During the ``tick'', agents perform their actions in parallel (AP). The environment is updated in EP fashion during ``tocks''. An efficient thread scheduler is used to balance the workload of both AP and EP stages of the simulation.

\section{Methodology}
\label{sec:methodology}

\subsection{Model overview}
\label{sec:mm:odd}

Here we present an overview of the PPHPC model. A complete description of the model (using the ODD protocol) is available in reference \cite{fachada2015template}.

\subsubsection{Purpose}
\label{sec:mm:odd:purpose}

The purpose of PPHPC is to serve as a standard model for studying and evaluating SABM implementation strategies. It is a realization of a predator\hyp{}prey dynamic system, and captures important characteristics of SABMs, such as agent movement and local agent interactions. The model can be implemented using substantially different approaches that ensure statistically equivalent results. Implementations may differ in aspects such as the selected system architecture, choice of programming language and/or agent\hyp{}based modeling framework, parallelization strategy, random number generator, and so forth. 

\subsubsection{Entities, state variables, scales}
\label{sec:mm:odd:ents}

The PPHPC model is composed of three entity classes: \textit{agents}, \textit{grid cells} and \textit{environment}. Each of these entity classes is defined by a set of state variables, as shown in Table~\ref{tab:ent:statevars}. Time\hyp{}dependent state variables are represented with uppercase letters, while constant state variables and parameters are denoted by lowercase letters.

\begin{table}[t!]
\centering
\begin{tabular}{llcl}
\toprule
Entity & State variable & Symbol & Range\\
\midrule
\multirow{8}{*}{Agents} 
    & Type                        & $t$   & $s, w$\\
    & Energy                      & $E$ & $1, 2, \ldots$ \\
    & Horizontal position in grid & $X$ & $0, 1, \ldots, x_{\text{env}}-1$\\
    & Vertical position in grid   & $Y$ & $0, 1, \ldots, y_{\text{env}}-1$\\
    & Energy gain from food       & $g^s$, $g^w$   & $0, 1, \ldots$\\
    & Energy loss per turn        & $l^s$, $l^w$   & $0, 1, \ldots$\\
    & Reproduction threshold      & $r_T^s$, $r_T^w$ & $1, 2, \ldots$\\
    & Reproduction probability    & $r_P^s$, $r_P^w$ & $0, 1, \ldots, 100$\\
\midrule
\multirow{3}{*}{Grid cells} 
    & Horizontal position in grid & $x$   & $0, 1, \ldots, x_{\text{env}}-1$\\
    & Vertical position in grid   & $y$   & $0, 1, \ldots, x_{\text{env}}-1$\\
    & Countdown                   & $C$ & $0, 1, \ldots, c_r$ \\
\midrule
\multirow{3}{*}{Environment} 
    & Horizontal size             & $x_{\text{env}}$   & $1, 2, \ldots$\\
    & Vertical size               & $y_{\text{env}}$   & $1, 2, \ldots$\\
    & Restart                     & $c_r$ & $1, 2, \ldots$\\
\bottomrule
\end{tabular}
\caption{\label{tab:ent:statevars}Model state variables by entity. Where applicable, the
$s$ and $w$ designations correspond to prey (\textit{sheep}) and predator (\textit{wolf}) 
agent types, respectively.}
\end{table}

The $t$ state variable defines the \textit{agent} type, either $s$ (\textit{sheep}, i.e. prey) or $w$ (\textit{wolf}, i.e. predator). The only behavioral difference between the two types is in the feeding pattern: while prey consume passive cell\hyp{}bound food, predators consume prey. Other than that, prey and predators may have different values for other state variables, as denoted by the superscripts $s$ and $w$. Agents have an energy state variable, $E$, which increases by $g^{\{s,w\}}$ when feeding, decreases by $l^{\{s,w\}}$ when moving, and decreases by half when reproducing. When energy reaches zero, the agent is removed from the simulation. Agents with energy higher than $r_T^{\{s,w\}}$ may reproduce with probability given by $r_P^{\{s,w\}}$. The grid position state variables, $X$ and $Y$, indicate which cell the agent is located in. There is no conceptual limit on the number of agents that can exist during the course of a simulation run. 

Instances of the \textit{grid cell} entity class can be thought of the place or neighborhood where agents act, namely where they try to feed and reproduce. Agents can only interact with other agents and resources located in the same grid cell. Grid cells have a fixed grid position, $(x, y)$, and contain only one resource, cell\hyp{}bound food (\textit{grass}), which can be consumed by prey, and is represented by the countdown state variable $C$. The $C$ state variable specifies the number of iterations left for the cell\hyp{}bound food to become available. Food becomes available when $C=0$, and when a prey consumes it, $C$ is set to $c_r$. 

The set of all grid cells forms the \textit{environment} entity, a toroidal square grid where the simulation takes place. The environment is defined by its size, $(x_{\text{env}},y_{\text{env}})$, and by the restart parameter, $c_r$.

Temporal extent is represented by a positive integer $m$, which denotes the number of discrete simulation steps or iterations.

\subsubsection{Process overview and scheduling}
\label{sec:mm:odd:procsched}

Algorithm~\ref{alg:jmain} describes the simulation schedule and its associated processes. Execution starts with an initialization process, \algfunc{Init()}. The process begins by instantiating the \textit{environment} entity and filling it with $x_{\text{env}} \times y_{\text{env}}$ grid cells. Cell\hyp{}bound food is initially available with 50\% probability. If not available, the countdown state variable, $C$, is set to a random value between $1$ and $c_r$. At this time, a predetermined number of agents are arbitrarily placed in the simulation environment, with initial energy set to a random value between 1 and $2g^{\{s,w\}}$.

\begin{algorithm}[t!]
\caption{Main simulation algorithm. \textbf{for} loops can be processed in \textit{any order} or in \textit{random order}. In terms of expected dynamic behavior, the former means the order is not relevant, while the latter specifies loop iterations should be explicitly shuffled.}
\label{alg:jmain}
\begin{algorithmic}[1]
\State \Call{Init}{\mbox{}}
\State \Call{GetStats}{\mbox{}}
\State $i\gets 1$
\For{$i<=m$}
	\ForAll{agent} 
    	\Comment{Any order}
    	\State \Call{Move}{\mbox{}}
    \EndFor
	\ForAll{grid cell} 
    	\Comment{Any order}
    	\State \Call{GrowFood}{\mbox{}}
    \EndFor
	\ForAll{agent} 
    	\Comment{Random order}
   		\State \Call{Act}{\mbox{}}
    \EndFor
	\State \Call{GetStats}{\mbox{}}
	\State $i\gets i+1$
\EndFor
\end{algorithmic}
\end{algorithm}

After initialization, and to get the simulation state at iteration zero, outputs are gathered by the \algfunc{GetStats()} process (discussed in further detail in section \ref{sec:mm:odd:design:obs}). The schedule then enters the main simulation loop, where each iteration is sub\hyp{}divided into four steps: 1) agent movement; 2) food growth in grid cells; 3) agent actions; and, 4) gathering of simulation outputs.

In step 1, agents \algfunc{Move()}, in any order, within a Von Neumann neighborhood, i.e. up, down, left, right or stay in the same cell, with equal probability. Agents lose $l^{\{s,w\}}$ units of energy when they move, even if they stay in the same cell; if energy reaches zero, the agent dies and is removed from the simulation. In step 2, during the \algfunc{GrowFood()} process, each grid cell checks if there is food available, i.e. if $C=0$. If not, i.e. if $C>0$, $C$ is decremented by one unit. In step 3, agents \algfunc{Act()} in explicitly random order, i.e. the agent list should be shuffled before the agents have a chance to act. During the \algfunc{Act()} process, agents first try to eat, and then try to reproduce.

\subsubsection{Parameterization}

Model parameters can be qualitatively separated into size\hyp{}related and dynamics\hyp{}related parameters, as shown in Table~\ref{tab:modparams}. Although size\hyp{}related parameters also influence model dynamics, this separation is useful for parameterizing simulations.

\begin{table}[t!]
\centering
\begin{tabular}{llc}
\toprule
Type & Parameter & Symbol\\
\midrule
\multirow{3}{*}{Size} 
    & Environment size          & $x_{\text{env}}, y_{\text{env}}$ \\
    & Initial agent count      & $P^s_0, P^w_0$ \\
    & Number of iterations     & $m$ \\
\midrule
\multirow{5}{*}{Dynamics} 
    & Energy gain from food    & $g^s$, $g^w$ \\
    & Energy loss per turn     & $l^s$, $l^w$ \\
    & Reproduction threshold   & $r_T^s$, $r_T^w$ \\
    & Reproduction probability & $r_P^s$, $r_P^w$ \\
    & Cell food restart        & $c_r$ \\
\bottomrule
\end{tabular}
\caption{\label{tab:modparams}Size\hyp{}related and dynamics\hyp{}related model parameters.}
\end{table}

Concerning size\hyp{}related parameters, more specifically, the grid size, we propose a base value of $100 \times 100$, associated with $400$ prey and $200$ predators. Different grid sizes should have proportionally assigned agent population sizes, as shown in Table~\ref{tab:initsize}.

\begin{table}[t!]
\centering
\begin{tabular}{rcrr}
\toprule
Size & $x_{\text{env}} \times y_{\text{env}}$ & $P^s_0$ & $P^w_0$ \\
\midrule
\num{100}  & $\num{100} \times \num{100}$   & \num{400}    & \num{200} \\
\num{200}  & $\num{200} \times \num{200}$   & \num{1600}   & \num{800} \\
\num{400}  & $\num{400} \times \num{400}$   & \num{6400}   & \num{3200} \\
\num{800}  & $\num{800} \times \num{800}$   & \num{25600}  & \num{12800} \\
\num{1600} & $\num{1600} \times \num{1600}$ & \num{102400} & \num{51200} \\
\hdashline[1.5pt/2pt]
\num{3200} & $\num{3200} \times \num{3200}$ & \num{409600} & \num{204800} \\
\num{6400} & $\num{6400} \times \num{6400}$ & \num{1638400} & \num{819200} \\
\num{12800} & $\num{12800} \times \num{12800}$ & \num{6553600} & \num{3276800} \\
$\vdots$ & $\vdots$ & $\vdots$  & $\vdots$ \\
\bottomrule
\end{tabular}
\caption{\label{tab:initsize}A selection of initial model sizes. Sizes above the dashed line are studied in detail in this work.}
\end{table}

For the dynamics\hyp{}related parameters, we propose two sets of parameters, Table~\ref{tab:paramsets}, which generate two distinct dynamics. The second parameter set typically yields more than twice the number of agents than the first parameter set. Matching results with runs based on distinct parameters is necessary in order to have a high degree of confidence in the similarity of different implementations \cite{edmonds2003replication}. While many more combinations of parameters can be experimented, these two sets are the basis which PPHPC implementations can use to check for correctness.

\begin{table}[t!]
\centering
\begin{tabular}{lcrr}
\toprule
Parameter & Symbol & Set 1 & Set 2 \\
\midrule
Prey energy gain from food & $g^s$ & 4 & 30 \\
Prey energy loss p/ turn  & $l^s$ & 1 & 1 \\
Prey reprod. threshold & $r_T^s$ & 2 & 2 \\
Prey reprod. probability  & $r_P^s$ & 4 & 10 \\
\midrule
Predator energy gain from food  & $g^w$ & 20 & 10 \\
Predator energy loss p/ turn & $l^w$ & 1 & 1 \\
Predator reprod. threshold & $r_T^w$ & 2 & 2 \\
Predator reprod. probability & $r_P^w$ & 5 & 5 \\
\midrule
Cell food restart & $c_r$ & 10 & 15 \\
\bottomrule
\end{tabular}
\caption{\label{tab:paramsets}Dynamics\hyp{}related parameter sets.}
\end{table}

While simulations of the PPHPC model are essentially non\hyp{}terminating\footnote{A non\hyp{}terminating simulation is one for which there is no natural event to specify the length of a run \cite{law2014simulation}.}, the number of iterations, $m$, is set to \num{4000}, as it allows to analyze steady\hyp{}state behavior for all the parameter combinations discussed here.

\subsubsection{Outputs} 
\label{sec:mm:odd:design:obs}

The following vector is collected in the \algfunc{GetStats()} process, where $i$ refers to the current iteration:

\begin{equation*}
\mathbf{O}_i = (P_i^s, P_i^w, P_i^c, \mean{E}_i^s, \mean{E}_i^w, \mean{C}_i)
\end{equation*}

$P^s_i$ and $P^w_i$ refer to the total prey and predator population counts, respectively, while $P^c_i$ holds the quantity of available cell\hyp{}bound food. $\mean{E}^s_i$ and $\mean{E}^w_i$ contain the mean energy of prey and predator populations. Finally, $\mean{C}_i$ refers to the mean value of the $C$ state variable in all grid cells.

\begin{figure}[t!]
\centering
\subfloat[Population, param. set 1.]{\includegraphics[width=0.48\linewidth]{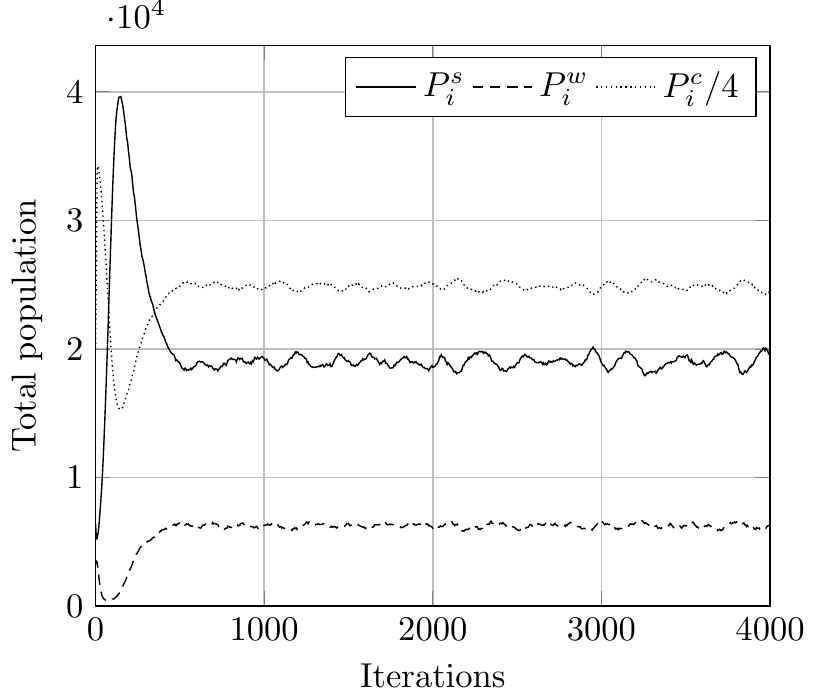}}\quad
\subfloat[Energy, param. set 1.]{\includegraphics[width=0.48\linewidth]{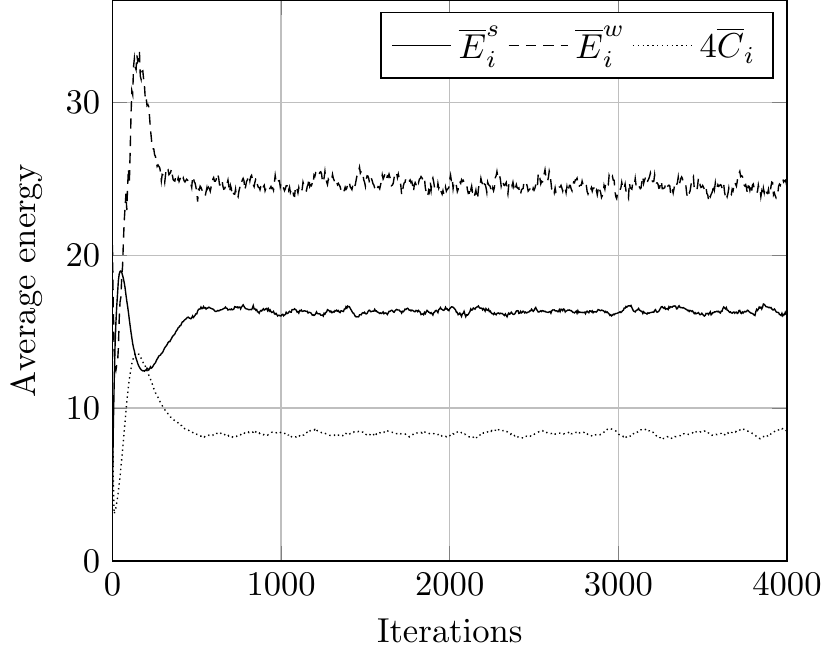}} \\
\subfloat[Population, param. set 2.]{\includegraphics[width=0.48\linewidth]{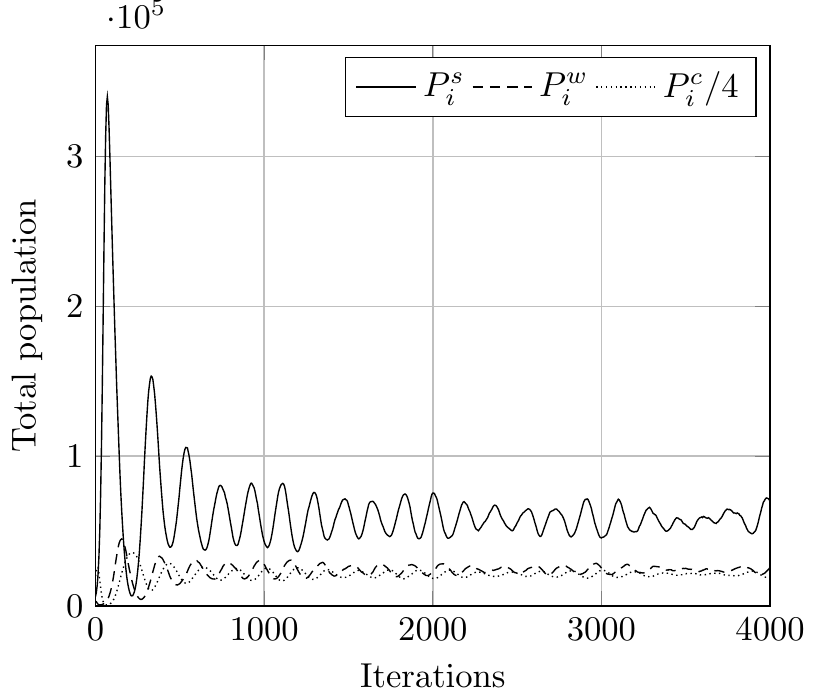}}\quad
\subfloat[Energy, param. set 2.]{\includegraphics[width=0.48\linewidth]{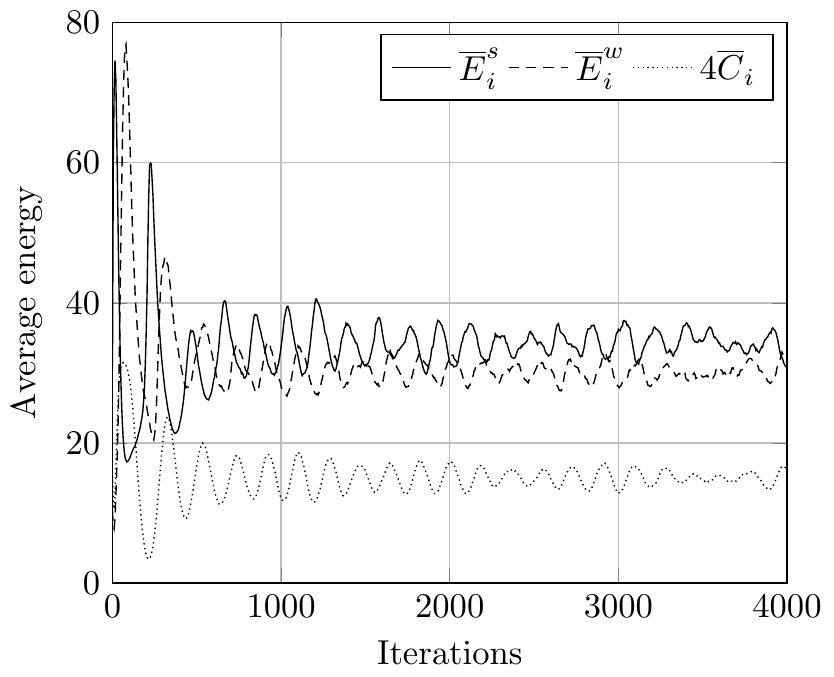}} \\
\caption{Typical model output for model size $400$. Other model sizes have similar outputs, apart from a vertical scaling factor. $P_i$ refers to total population, $\mean{E}_i$ to mean energy and $\mean{C}_i$ to mean value of the countdown state variable, $C$. Superscript $s$ relates to prey, $w$ to predators, and $c$ to cell\hyp{}bound food. $P_i^c$ and $\mean{C}_i$ are scaled for presentation purposes.}
\label{fig:typicalmod}
\end{figure}

Figure \ref{fig:typicalmod} shows the typical output of a simulation run with size 400 for both parameter sets. Model sizes up to \num{12800} have similar outputs, apart from a vertical scaling factor.

\subsection{A multithreaded Java implementation}
\label{sec:mm:impl}

Java is a general\hyp{}purpose, object\hyp{}oriented computer programming language, and is usually compiled to bytecode that runs on a Java Virtual Machine (JVM) \cite{gosling2014}. Version 5.0 of the Java programming language marked a huge step forward for the development of concurrent applications, offering new higher\hyp{}level components and additional low\hyp{}level mechanisms that make it simpler to build concurrent applications \cite{goetz2006java}. Java is a well\hyp{}known language within the ABM community, powering popular toolkits such as Repast Simphony \cite{north2013complex} and MASON \cite{luke2005mason}. Additionally, NetLogo also runs on the JVM, making a Java implementation of PPHPC even more appropriate for performance comparison purposes.

The Java implementation of PPHPC is based on the concept of units of work, which are processed by one or more worker threads. 
The basic unit of work is a single grid cell, except in the agent initialization stage (which takes place during the \algfunc{Init()} process), where the unit of work corresponds to the instantiation and deployment of a single agent. 

Work providers supply worker threads with tokens uniquely identifying the units of work to be processed. Different work providers offer specific parallelization strategies, as will be discussed in section \ref{subsubsec:java:worktypes}.

\subsubsection{Architecture}
\label{subsubsec:java:arch}

\begin{figure}[t!]
\centering
\includegraphics[width=\linewidth]{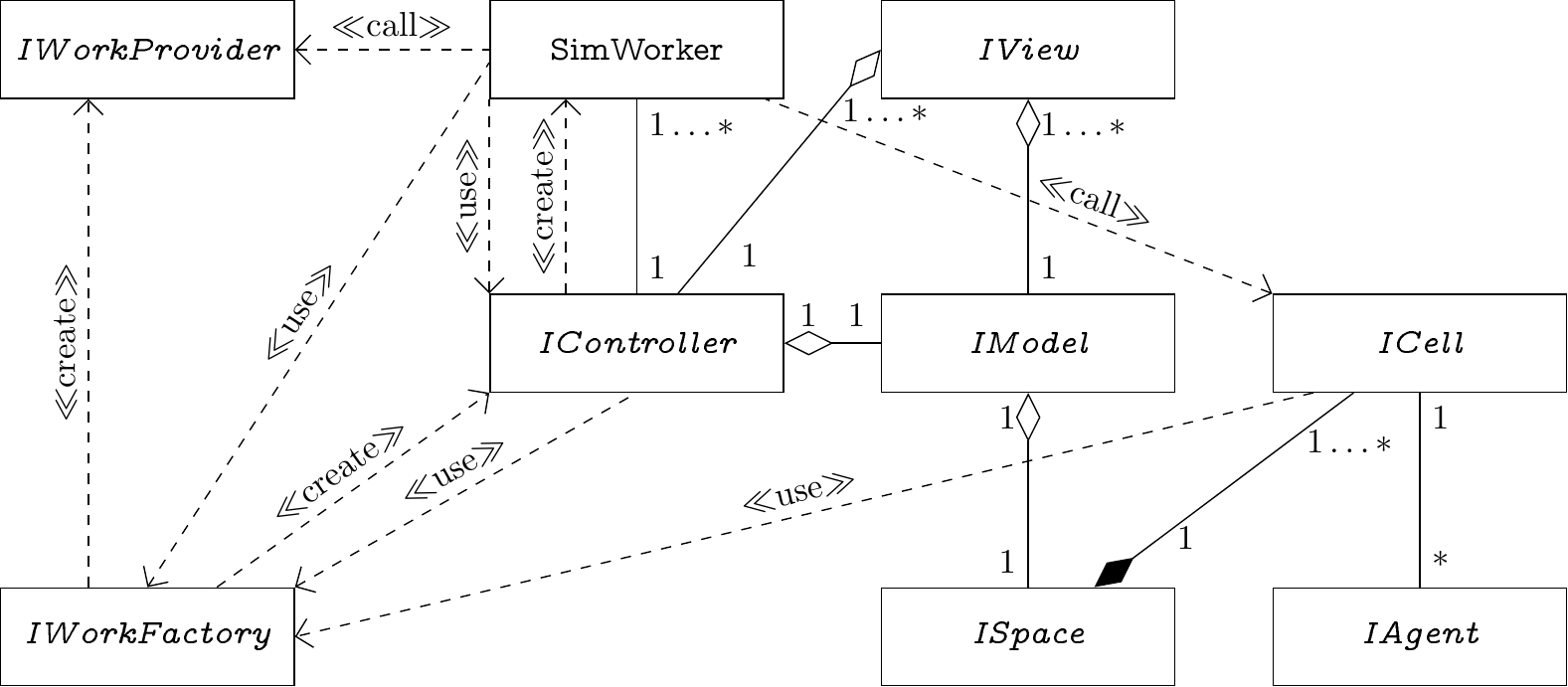}
\caption{UML diagram for the multithreaded Java implementation of PPHPC.}
\label{fig:juml}
\end{figure}

The Java implementation is built upon the Model\hyp{}View\hyp{}Controller design pattern \cite{gamma1994design}, as shown in Figure \ref{fig:juml}. The model (generically represented by the \javainterface{IModel} interface) contains the actual ABM logic, aggregating the simulation grid (interface \javainterface{ISpace}), composed of grid cells (interface \javainterface{ICell}). Cells are associated with zero or more agents (interface \javainterface{IAgent}). The model can be manipulated and observed using one or more views, represented by the \javainterface{IView} interface. Views observe the model directly but manipulate it (e.g. start, pause, stop) via the controller (interface \javainterface{IController}). Work factories (classes implementing the \javainterface{IWorkFactory} interface), are responsible for creating objects which handle how units of work are processed, namely the controller and the work provider (the latter represented by the \javainterface{IWorkProvider} interface).

The controller spawns a specified number of worker threads (instances of the \javaclass{SimWorker} class), which will execute Algorithm \ref{alg:jmain}. The quantity of work each worker processes is determined by the work provider, i.e. by the tokens it provides. To improve performance, workers execute the operations in Algorithm \ref{alg:jmain} in a different order, but in a way that the final qualitative simulation outcome does not change, as outlined in Algorithm \ref{alg:jsimwork}\footnote{Controller synchronization points (i.e. calls to \algfunc{ControllerSync()}) are discussed in section \ref{subsubsec:java:worktypes}.}. 

\begin{algorithm}[t!]
\caption{Realization of the main simulation algorithm in the \javaclass{SimWorker} class. Calls to \algfunc{ControllerSync()} are controller synchronization points.}
\label{alg:jsimwork}
\begin{algorithmic}[1]
\State \Call{ControllerSync}{1} 
\State \Call{CreateCells}{\mbox{}} \Comment Environment\hyp{}parallel 
\State \Call{ControllerSync}{2} 
\State \Call{SetCellNeighbors}{\mbox{}} \Comment Environment\hyp{}parallel
\State \Call{ControllerSync}{3} 
\State \Call{CreatePrey}{\mbox{}} \Comment Agent\hyp{}parallel 
\State \Call{CreatePredators}{\mbox{}} \Comment Agent\hyp{}parallel 
\State \Call{ControllerSync}{4} 
\State \Call{GetStats}{\mbox{}} \Comment Environment\hyp{}parallel
\State \Call{ControllerSync}{5} 
\State $i\gets 1$
\For{$i<=m$}
	\State \Call{Move}{\mbox{}} + \Call{GrowFood}{\mbox{}} \Comment Environment\hyp{}parallel
	\State \Call{ControllerSync}{6} 
	\State \Call{Act}{\mbox{}} + \Call{GetStats}{\mbox{}} \Comment Environment\hyp{}parallel
	\State \Call{ControllerSync}{7} 
	\State $i\gets i+1$
\EndFor
\State \Call{ControllerSync}{8} 
\end{algorithmic}
\end{algorithm}

The \algfunc{Init()} process, line 1 of Algorithm \ref{alg:jmain}, is divided in four steps, which can be processed in parallel, as determined by the work provider: 1) instantiation and initialization of grid cells (line 2 of Algorithm \ref{alg:jsimwork}); 2) connecting cells to their neighbors (line 4 of Algorithm \ref{alg:jsimwork}); 3) instantiation of prey (line 6 of Algorithm \ref{alg:jsimwork}); and, 4) instantiation of predators (line 7 of Algorithm \ref{alg:jsimwork}). In steps 1 and 2 the unit of work represents a single grid cell. In steps 3 and 4 the unit of work represents the instantiation and deployment of a single agent.

Next, workers execute the \algfunc{GetStats()} process (line 2 of Algorithm \ref{alg:jmain} and line 9 of Algorithm \ref{alg:jsimwork}). Data required for the observation of the model, specified in section \ref{sec:mm:odd:design:obs}, is collected on a cell\hyp{}by\hyp{}cell basis. After processing their allocated cells, workers then update a global statistics object.

Lines 5 to 10 of Algorithm \ref{alg:jmain}, which include agent movement and growth of cell\hyp{}bound food, are condensed into a single EP loop (line 13 of Algorithm \ref{alg:jsimwork}). This is possible because the \algfunc{Move()} and \algfunc{GrowFood()} processes are independent; i.e. the consequences of either will only impact the \algfunc{Act()} process, which occurs later. Most importantly, both \algfunc{Move()} and \algfunc{GrowFood()} are cell\hyp{}wise independent and can be processed autonomously for each cell in an EP loop. When a cell is processed, agents located therein are prompted to move, and then the cell is asked to execute its \algfunc{GrowFood()} process. Care is taken so that agents that already moved are not prompted to move again.

Finally, lines 11 to 14 of Algorithm \ref{alg:jmain}, containing the \algfunc{Act()} and \algfunc{GetStats()} processes, are also contracted into one EP loop, as shown in line 15 of Algorithm \ref{alg:jsimwork}. While agent actions and statistics gathering are not independent events (i.e. the former must occur before end\hyp{}of\hyp{}iteration data is obtained), they are cell\hyp{}wise independent. As specified in section \ref{sec:mm:odd:ents}, the actions of an agent are limited to the cell it occupies. Thus, after agent actions take place in a cell, end\hyp{}of\hyp{}iteration cell data is collected. Note that before agents in a cell act, the local agent list is shuffled according to Algorithm \ref{alg:jmain}, i.e. \textit{random order}. After processing all of their allocated cells, workers then update a global statistics object.

Lines 13 and 15 of Algorithm \ref{alg:jsimwork} reflect the separation of each simulation step into two parts, in a ``tick\hyp{}tock'' fashion. This is often necessary in multithreaded ABM implementations to synchronize agent movement, agent actions, environmental dynamics or reading and writing of shared environment areas \cite{wei2013efficient,goldsby2013multithreaded}. 

\subsubsection{Random number generators and reproducible simulations}
\label{subsubsec:java:rng}

Reproducibility of simulations is an important and often overlooked aspect of ABM parallelization. As explained by Hill et al. \cite{hill2013distribution}, ``To investigate and understand the results, we have to reproduce the same scenarios and find the same confidence intervals every time we run the same stochastic experiment. When debugging parallel stochastic applications, we need to reproduce the same control flow and the same result to correct an anomalous behavior''. Most ABMs, including the one presented here, use one or more stochastic processes.

PRNGs use iterative deterministic algorithms for producing a sequence of pseudo\hyp{}random numbers that approximate a truly random sequence \cite{coddington1997random}. A PRNG consists of a finite set of states, and a transition function that takes the PRNG from one state to the next. The initial state of the PRNG is called the seed \cite{srinivasan2003testing}. As such, PRNGs are used in simulations to mimic stochastic processes in a reproducible fashion.

Reproducibility is simple to accomplish in a single\hyp{}threaded implementation, sufficing the use of the same PRNG and seed, with deterministic scheduling of all stochastic processes. However, in a parallel simulation context, there are practical trade\hyp{}offs between reproducibility, memory and speed \cite{voss2010scalable}. Additionally, the use of PRNGs in parallel simulations comes with its own set of problems, such as hidden correlations or overlaps in different sub\hyp{}streams of the same PRNG \cite{hill2013distribution}.

For the Java PPHPC implementation, parallel pseudo\hyp{}random generation is carried out in the following manner. The $i^{\text{th}}$ worker thread obtains its own sub\hyp{}sequence of a global random sequence by using a unique seed, $S_i$, through a random spacing approach \cite{hill2013distribution}. Each seed $S_i$ is derived from the worker's unique identifier, $i$, and from a user specified global seed, $S$, according to Eq. \ref{eq:jrngseed},

\begin{equation}
\label{eq:jrngseed}
S_i=
\begin{cases}
    S,                         & \text{if } i=0 \\
    S \oplus \text{SHA256}(i), & \text{if } i>0
\end{cases}
\end{equation}

\noindent where $\oplus$ is the bitwise XOR operator and $\text{SHA256}()$ is the SHA-256 cryptographic hash function. The Java implementation of PPHPC uses the Uncommons Maths library \cite{dyer2012}, taking advantage of the several PRNGs it provides. For the results presented in this paper, the library's Mersenne Twister implementation was used, because it is the same PRNG used by NetLogo, and its very large period of $2^{19937}-1$ makes sub\hyp{}stream overlapping highly unlikely to occur \cite{brugger2014parallel,steele2014fast}.

Considering that each worker has its own PRNG sub\hyp{}sequence, the following conditions are required in order to make simulations reproducible:

\begin{enumerate}
\item Each worker must process the exact same work between runs, i.e. it must:
	\begin{enumerate}
		\item Instantiate the same quantity of initial agents and place them in the same cells.
		\item Process the same cells.
	\end{enumerate}
\item Agents within a cell must be processed in the same order.
\end{enumerate}

The first condition can be guaranteed if work providers always assign the same tokens to the same worker. The second condition may be problematic when cell\hyp{}level synchronization is required, which occurs during agent movement. As will be discussed in the next section, this issue can be solved within the current framework by placing agents in their destination cell using some deterministic order criteria. This ensures that, when entering the agent action stage, each cell contains an ordered list of agents, ready to act. 

A broader way of forcing reproducibility of simulations is to associate the PRNG sub\hyp{}sequences with cells and not with worker threads. However, this can result in a large number of parallel PRNG states (e.g. $1.6 \times 10^5$ PRNG states are required for model size $400$). There are two main problems with this approach. The first, already briefly discussed, is that the problems associated with parallel PRNGs become worse when partitioning the PRNG stream into more and more sub\hyp{}streams \cite{hill2013distribution}. The second issue is related to the required memory. For the problem of size $400$, using a large period PRNG (in order to minimize the first problem) such as the Mersenne Twister, will approximately require 380 MiB of memory just for the PRNG states; for larger model sizes, such as $\num{1600}$, almost 6 GiB are required. 

\subsubsection{Parallelization strategies}
\label{subsubsec:java:worktypes}

A parallelization strategy is defined by the selected work factory, more specifically by the work providers it offers, and by the way it configures the controller. The work factory is first requested to instantiate and configure an appropriate controller object. When the simulation starts, the controller spawns the number of workers specified by the work factory, and each worker gets a reference to the controller and to the work factory. Workers use the former to synchronize themselves and the latter to get references to work providers; these, in turn, provide workers with tokens, i.e. integers that uniquely identify units of work. When a work provider returns a negative value, it means that the current parallel work cycle is finished. Three work providers are used by workers: one for the cells, which provides work in an EP fashion for all cell\hyp{}wise work cycles, and two for the initial agent creation (one for prey, the other for predators), which provide AP work. The latter are used only once during the \algfunc{Init()} process, while the former, i.e. the cell work provider, is continuously reused.

To accommodate different parallelization strategies, workers have several possible synchronization points, which occur at three different levels: 1) controller; 2) work provider; and 3) grid cells.

There are eight controller\hyp{}level synchronization points. All workers explicitly notify the controller when they reach them, as shown by the calls to \algfunc{ControllerSync()} in Algorithm \ref{alg:jsimwork}. Whether or not workers are held on that point by the remaining workers depends on how the controller was configured by the work factory. There are, however, some points where barriers are mandatory. For example, no worker can begin processing agent actions before all agents have moved and all food has grown; thus, synchronization point 6 (line 14 of Algorithm \ref{alg:jsimwork}) is necessarily a barrier. Sequences of concurrent computation (AP or EP) and thread communication, terminating with a barrier, can be considered global \textit{supersteps} under a Bulk Synchronous Parallel model \cite{valiant1990bridging}.

Work provider and cell\hyp{}level synchronization is performed implicitly when workers request work and when they process cells, respectively. There are two possible cell\hyp{}level synchronization points: 1) when inserting initial agents during the \algfunc{Init()} process; and, 2) when inserting agents moving from other cells. Again, whether or not workers are actually synchronized at work provider and cell\hyp{}level synchronization points depends on the parallelization strategy. Five parallelization strategies are provided, as shown in Table~\ref{tab:worktypes}, which also enumerates all possible synchronization points and how the different strategies handle them. The following paragraphs describe in detail these parallelization strategies.

\begin{table}[!t]
\centering
\begin{tabular}{cccccccccccc}
\toprule
\multirow{2}{*}{PS} & \multicolumn{8}{c}{Controller} & \multirow{2}{*}{WP} & \multicolumn{2}{c}{Cell}\\
\cmidrule(r){2-9} \cmidrule(r){11-12}
 & 1 & 2 & 3 & 4 & 5 & 6 & 7 & 8 &  & Init. & Move \\  
\midrule
ST & -   & -   & -   & -   & -   & -   & -   & -   & -   & -         & - \\
EQ & $S$ & $B$ & $S$ & $B$ & $S$ & $B$ & $B$ & $S$ & -   & $S$       & $S$ \\
EX & $S$ & $B$ & $S$ & $B$ & $S$ & $B$ & $B$ & $S$ & -   & $\widehat{S}$ & $\widehat{S}$\\
ER & $S$ & $B$ & $S$ & $B$ & $S$ & $B$ & $B$ & $S$ & $B$ & $\widehat{S}$ & - \\
OD & $S$ & $B$ & $S$ & $B$ & $B$ & $B$ & $B$ & $S$ & $S$ & $S$       & $S$ \\
\bottomrule
\end{tabular}
\caption{\label{tab:worktypes}Parallelization strategies (PS) and their handling of the possible synchronization points at the level of the controller, work provider (WP) and grid cell. $B$ means there is a barrier, i.e. that workers can only advance when all workers have reached the sync. point. $S$ implies access serialization, but workers do not have to wait on other workers before continuing; $\widehat{S}$ is similar, but implies an ordered agent insertion, which might take longer than simple access serialization.}
\end{table}

\paragraph{Single\hyp{}thread (ST)}

A single\hyp{}threaded work factory (\javaclass{SingleThreadWorkFactory} class) is provided for comparison with the multithreaded work factories. This work factory configures the controller such that no synchronization occurs when the (single) worker explicitly notifies the controller that it reached a given synchronization point. Likewise, no work provider or cell\hyp{}level synchronization is required. The work providers made available by the single\hyp{}threaded work factory (instances of the \javaclass{SingleThreadWorkProvider} class) maintain a simple counter which issues tokens to the single worker. Simulation objects (cells and agents) are always iterated in the same order, which allows for simulation reproducibility. 

\paragraph{Equal (EQ)}

The general idea of the EQ parallelization strategy (handled by the \javaclass{EqualWorkFactory} class) is that each worker always processes the same work. Work distribution is performed once at the beginning of the simulation by the associated work providers (instances of the \javaclass{EqualWorkProvider} class), and then the workers are always given the same exact tokens, e.g. they always process the same cells in the EP sections. Cell\hyp{}level synchronization is required because more than one worker may potentially access the same cell at the same time for agent movement or initial agent placement. The first worker to whom access is granted gets to place its agent topmost in the cell's internal agent list. This means that, because thread synchronization is not a deterministic process, agents will not be processed in the same order from run to run. Thus, simulations with this work factory are not reproducible. The maximum number of tokens to be processed by each worker, $n$, is given by

\begin{equation}
\label{eq:jeq_workdiv_tpw}
n=\lceil T/N \rceil
\end{equation}

\noindent where $T$ is the number of tokens to be processed in a parallel work cycle, and $N$ is the number of worker threads. If $T$ is not equally divisible between the available workers, the last worker will process less work than the remaining workers, as shown in Eq. \ref{eq:jeq_workdiv_t},

\begin{equation}
\label{eq:jeq_workdiv_t}
n \cdot i \leq t_i < \min \big( n \cdot (i+1),T \big)
\end{equation}

\noindent where $i$ identifies the $i^{\text{th}}$ worker, and $t_i$ corresponds to the range of tokens which will be processed by the $i^{\text{th}}$ worker.

\paragraph{Equal with repeatability (EX)}

The EX parallelization strategy is a slight variation on EQ. The same classes are used, the only difference is that when workers access the same cell at the same time for agent placement purposes, the agent is placed in ordered fashion, as shown in Table~\ref{tab:worktypes}. This allows for reproducible simulations because agents will be processed in the same order. 

\paragraph{Equal with row synchronization (ER)}

\begin{figure}[t!]
\centering
  \includegraphics[width=\linewidth]{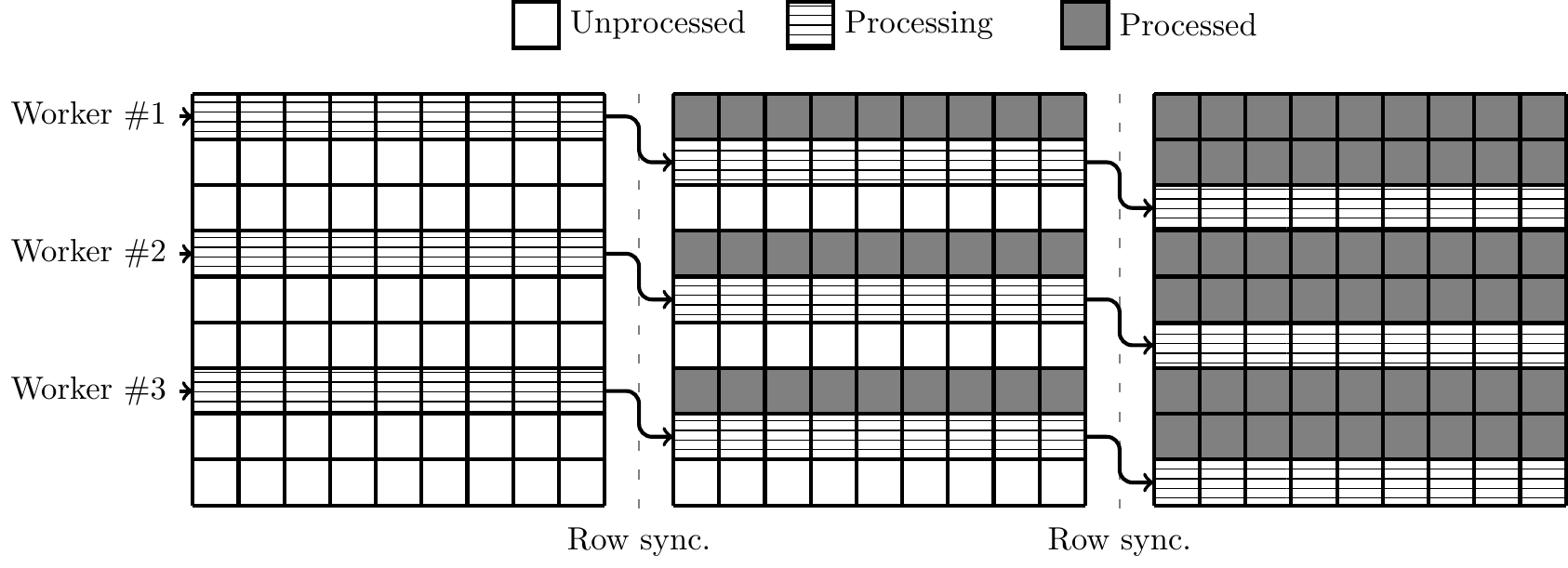}
  \caption{Equal with row synchronization (ER): example of three workers processing nine rows of the simulation grid in parallel.}
  \label{fig:jer}
\end{figure}

In the ER parallelization strategy, as in the previous strategies, work is assigned to workers at the beginning of the simulation. The difference here is that each worker serially processes rows of the simulation grid, leaving a distance of at least three rows (including the row to be processed) to the next worker (see Figure \ref{fig:jer}). More generally, this distance is given by

\begin{equation}
\label{eq:jer_dmin}
d_{\text{min}}=2r+1
\end{equation}

\noindent where $r$ is the agent movement radius, which is 1 for the PPHPC model. This approach allows workers to run in parallel without any need for cell\hyp{}level synchronization, because they synchronize at the end of each row at the work provider level. Thus, agents always move to neighboring cells in the same order, making simulations reproducible.

The \javaclass{EqualRowSyncWorkFactory} class is responsible for configuring the controller and issuing the appropriate work providers. An instance of the \javaclass{EqualRowSyncWorkProvider} class is issued for EP work. However, this approach does not make sense for AP work; as such, the agent initialization phase, which is handled in AP fashion, is managed by instances of the \javaclass{EqualWorkProvider} class. For simulations to be reproducible, initial agents are inserted in an ordered fashion, as shown in Table~\ref{tab:worktypes}.

This strategy implies that there is a practical maximum number of workers depending on the number of rows, $y_{\text{env}}$, and on the minimum distance between rows, $d_{\text{min}}$, as shown in Eq. \ref{eq:jer_Nmax}:

\begin{equation}
\label{eq:jer_Nmax}
N_{\text{max}}=\left\lfloor \frac{y_{\text{env}}}{d_{\text{min}}} \right\rfloor
\end{equation}

\noindent If the specified number of workers, $N$, is larger than $N_{\text{max}}$, an exception is thrown and the simulation terminates. An initial estimate of the number of rows per worker, $\Delta y$, is given by

\begin{equation}
\label{eq:jer_rpw}
\Delta y=\left\lfloor \frac{y_{\text{env}}}{N} \right\rfloor
\end{equation}

This estimate can be incremented if: 1) the number of rows is not equally divisible by the number of workers; and, 2) after incrementing it, there are enough rows for the last worker to process. This is shown in Eq. \ref{eq:jer_rpwf}, where $\Delta y_{f}$ is the final number of rows per worker:

\begin{equation}
\label{eq:jer_rpwf}
\Delta y_{f}=
\begin{cases}
    \Delta y + 1, & \text{if } y_{\text{env}} \text{ mod } N > 0 \, \land \, (N-1) \cdot (\Delta y + 1) \leq y_{\text{env}} - d_{\text{min}} \\
    \Delta y,     & \text{otherwise.}
\end{cases}
\end{equation}

From the workers perspective, what matters are the tokens to process. All workers, except possibly the last one, will process $n$ tokens, according to Eq. \ref{eq:jer_n}. The exact tokens that the $i^{\text{th}}$ worker will process are given in Eq. \ref{eq:jer_workdiv_t}. Note that any adjustment due to the number of rows not being exactly divisible by the number of workers is performed on the last worker.

\begin{equation}
\label{eq:jer_n}
n=x_{\text{env}} \cdot \Delta y_{f}
\end{equation}

\begin{equation}
\label{eq:jer_workdiv_t}
n \cdot i \leq t_i < t_f \text{, where } t_f=
\begin{cases}
	n \cdot (i+1), & \text{if } i<N-1 \\
	x_{\text{env}} \times y_{\text{env}}, & \text{if } i=N-1
\end{cases}
\end{equation}

\paragraph{On\hyp{}demand (OD)}

The OD parallelization strategy, managed by an instance of the \javaclass{OnDemandWorkFactory} class, aims to improve load balancing by issuing smaller blocks of tokens to keep workers busy. Work providers, instances of the \javaclass{OnDemandWorkProvider} class, maintain a counter of the tokens already issued to workers. Each time a worker requests more tokens, this counter is incremented by the block size, $b$. Access to the work provider is serialized, as shown in Table~\ref{tab:worktypes}, because the work counter needs to be atomically incremented. The counter is implemented using an instance of the \javaclass{AtomicInteger} class, which was added to the Java SE 5.0 API. Lower values of $b$ will cause workers to fetch tokens from the work provider more often, which may cause some thread contention; however, work distribution is improved because workers are more likely to be processing work instead of waiting for slower workers at controller synchronization points. Conversely, with higher values of $b$, worker threads will request work less frequently, which leads to lower thread contention; on the downside, faster workers may have to wait longer for slower ones. Thus, $b$ controls a trade\hyp{}off between thread contention and load balancing. If $b$ is selected such that workers only access the work provider once and the same amount of work is assigned to each worker, the OD strategy becomes similar to EQ, but with additional synchronization. 

Table \ref{tab:worktypes} also shows that the OD strategy requires all workers to explicitly wait for one another at controller synchronization point 5. This is required because some workers may finish processing their \algfunc{GetStats()} tokens early (line 9 of Algorithm~\ref{alg:jsimwork}), while others lag behind; at line 13 of Algorithm~\ref{alg:jsimwork}, faster workers could potentially obtain work tokens that are still being processed by slower workers at line 9. 

This parallelization strategy does not offer reproducible simulations because workers obtain tokens in a FIFO fashion which is dependent on the OS thread scheduling, and thus not deterministic. As such, it is not possible to anticipate which worker will process which tokens, resulting in cells being associated with different worker\hyp{}bound PRNG sub\hyp{}sequences from iteration to iteration and from run to run.

\subsubsection{Source code}

The Java implementation described here is available at \url{https://github.com/fakenmc/pphpc/tree/java/java}.

\subsection{Comparison of multiple implementations} 
\label{sec:mm:compare}


Axtell et al. \cite{axtell1996aligning} define three replication or comparison standards for the level of similarity between model outputs: \textit{numerical identity}, \textit{distributional equivalence} and \textit{relational alignment}. The first, \textit{numerical identity}, implies exact numerical output, but it is difficult to attain and not critical for showing that two models have the same type of dynamic behavior. To achieve this goal, \textit{distributional equivalence} is a more appropriate choice when parallelizing ABMs, as it aims to reveal the statistical similarity between two outputs. 
Finally, \textit{relational alignment} between two outputs exists if they show qualitatively similar dependencies with input data, which is frequently the only way to compare a model with another which is inaccessible (e.g. implementation has not been made available by the original author), or with a non\hyp{}controllable ``real'' system (such as a model of the human immune system \cite{NunoFachada2008}).

Given that we are essentially interested in showing that the parallel variants of the Java implementation have statistically indistinguishable behavior from the canonical NetLogo implementation, we aim for the comparison standard of distributional equivalence. In order to compare multiple versions of PPHPC for distributional equivalence, we follow the approach taken by Wilensky and Rand \cite{wilensky2007making}, demonstrating the distributional equivalence of a few statistical summaries for each output, avoiding the need of trying to show distributional equivalence for every output at every iteration. This process can be realized with the following steps: 1) select a set of statistics which summarize each output (individual statistics from each output are designated as focal measures); 2) for each version of PPHPC or parallelization strategy, obtain samples of individual focal measures from a number of replications (each replication performed with a different PRNG seed); and, 3) apply a statistical test to the several samples of individual focal measures to check if they are distributionally equivalent.

In the first step we select, for each of the six outputs, the six statistical summaries described in Table~\ref{tab:focmeasures}, in a total of 36 focal measures. The rationale behind this selection, including the choice of $l$, which separates the transient and steady\hyp{}state stages of the simulation, is discuss in detail in reference \cite{fachada2015template}. 

\begin{table}[t!]
\centering
\begin{tabular}{ll}
\toprule
Statistic & Description\\
\midrule

$\max\limits_{0 \leq i \leq m}{X_i}$ & Maximum value. \\
$\argmax\limits_{0 \leq i \leq m}{X_i}$  & Iteration where maximum value occurs. \\
$\min\limits_{0 \leq i \leq m}{X_i}$ & Minimum value. \\
$\argmin\limits_{0 \leq i \leq m}{X_i}$ & Iteration where minimum value occurs. \\
$\mean{X}^{\text{ss}}=\sum^{m}_{i=l+1}X_i/(m-l)$ & Steady\hyp{}state mean. \\
$S^{\text{ss}}=\sqrt{\dfrac{\sum^{m}_{i=l+1}(X_i-\mean{X}_{\text{ss}})^2}{m-l-1}}$ & Steady\hyp{}state sample standard deviation. \\

\bottomrule
\end{tabular}
\caption{\label{tab:focmeasures}Statistical summaries for each output $X$, where $X_i$ is the value of $X$ at iteration $i$, $m$ denotes the last iteration, and $l$ corresponds to the iteration separating the transient and steady\hyp{}state stages. For parameter set 1, $l=1000$, while for parameter set 2, $l=2000$.}
\end{table}

For step 2, a number of replications was performed for each version of PPHPC or parallelization strategy, with the global PRNG seed, $S$, taken from the MD5 checksum of the replication number such that individual replications are performed with essentially uncorrelated seeds.

Concerning step 3, a number of statistical tests can be used to determine if focal measures from different PPHPC versions or parallelization strategies are distributionally equivalent. The choice of statistical test(s) depends on a number of factors, such as the number of samples to be compared or whether a specific focal measure follows a normal distribution. Regarding the former, there are six samples to be compared for each focal measure, produced by the NetLogo implementation and the five parallel variants of the Java implementation. Concerning the latter, some focal measures follow an approximately normal distribution, while others do not \cite{fachada2015template}. Consequently, a statistical test which is able to simultaneously compare six samples and does not assume normality of sample populations is desirable. The Kruskal\hyp{}Wallis test \cite{kruskal1952use} fills these requirements. It is the non\hyp{}parametric equivalent of ANOVA, allowing us to compare samples drawn from populations with normal and non\hyp{}normal distributions, extending the Mann\hyp{}Whitney U test for more than two samples \cite{gibbons2011nonparametric}. It tests the null hypothesis that all samples are drawn from the same distribution by comparing their medians. As such, it is used in this work to compare the outputs of the six PPHPC realizations.

\section{Results and Discussion}
\label{sec:results}

A total of 10 replications were performed with the following combination of parameters:

\begin{itemize}
	\item \emph{Implementations (variants)}: NetLogo (NL) and Java (ST, EQ, EX, ER and OD)
	\item \emph{Parameter sets}: 1, 2
	\item \emph{Model sizes}: 100, 200, 400, 800, 1600
	\item \emph{Number of threads (multithreaded Java variants only)}: 1, 2, 4, 6, 8, 12, 16, 24
	\item \emph{Block size (Java OD variant only)}: 20, 50, 100, 200, 500, 1000, 2000, 5000
\end{itemize}

This implies a total of \num{20} and \num{1780} replications per size for the NetLogo and Java implementations, respectively, causing practical difficulties in scaling the experiment to sizes above \num{1600}. For each combination of parameters, the 10 replications were performed with distinct PRNG seeds. These replications are the basis for both the performance and statistical analysis performed in the next sections. All performance results are based on the mean run time of the 10 replications. 

Replications were performed in ``headless'' mode for both implementations, i.e., without any graphical component. For the NetLogo implementation this meant the use of the BehaviorSpace tool from the command line. Parallel runs were disabled because we were interested in benchmarking the performance of individual runs, and simultaneous runs may have interfered in the measurements. Replications with the Java implementation were performed with a single non\hyp{}GUI MVC view, which performs a simulation from start to finish without user intervention.

All replications were performed on a machine with the following hardware and software configuration:

\begin{itemize}
\item Intel(R) Core(TM) i7\hyp{}3930K CPU 3.20GHz (six cores, two logical processors per core), 32GB RAM
\item Ubuntu $14.04.2$ LTS, OpenJDK Java $1.7.0$, NetLogo $5.1.0$
\end{itemize}

The data produced by this computational experiment, as well as the scripts used to set up the experiment, are made available to other researchers at \url{https://zenodo.org/record/34049}.

\subsection{Performance comparison}
\label{subsec:perf}

\begin{figure}[!t]
\centering
\subfloat[Speedup against NL, param. set 1.\label{fig:speedups:nlv1}]{\includegraphics[width=0.48\linewidth]{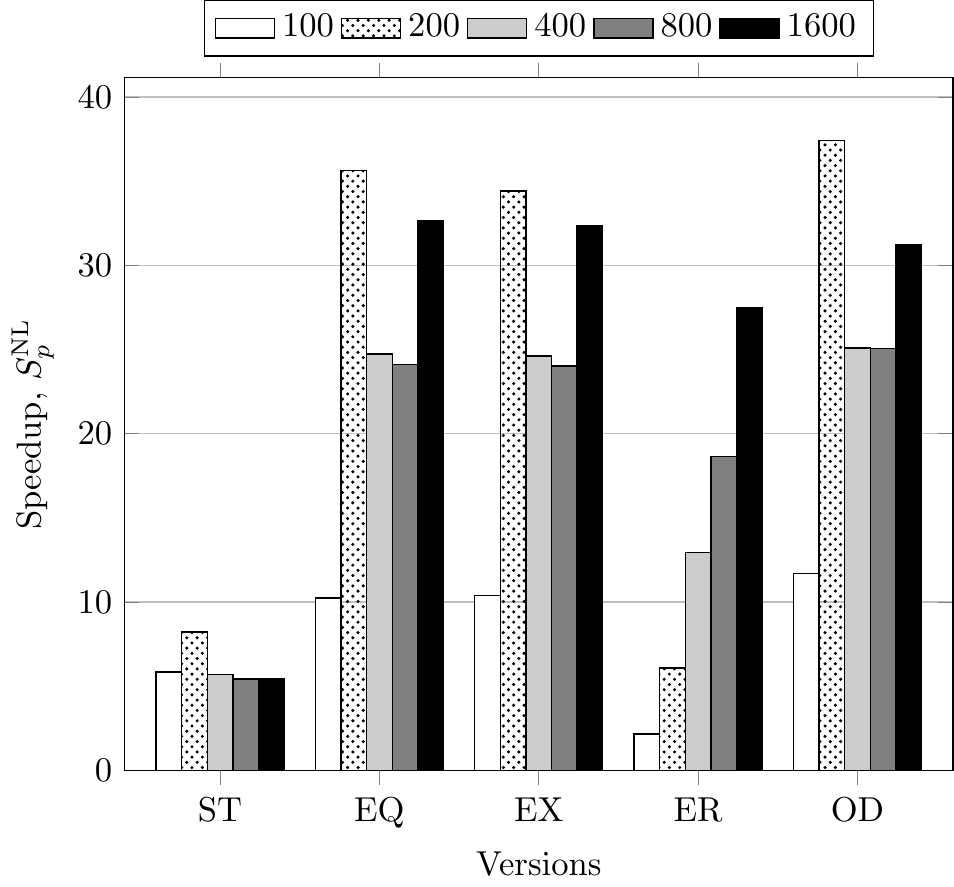}}\quad
\subfloat[Speedup against ST, param. set 1.\label{fig:speedups:jv1}]{\includegraphics[width=0.48\linewidth]{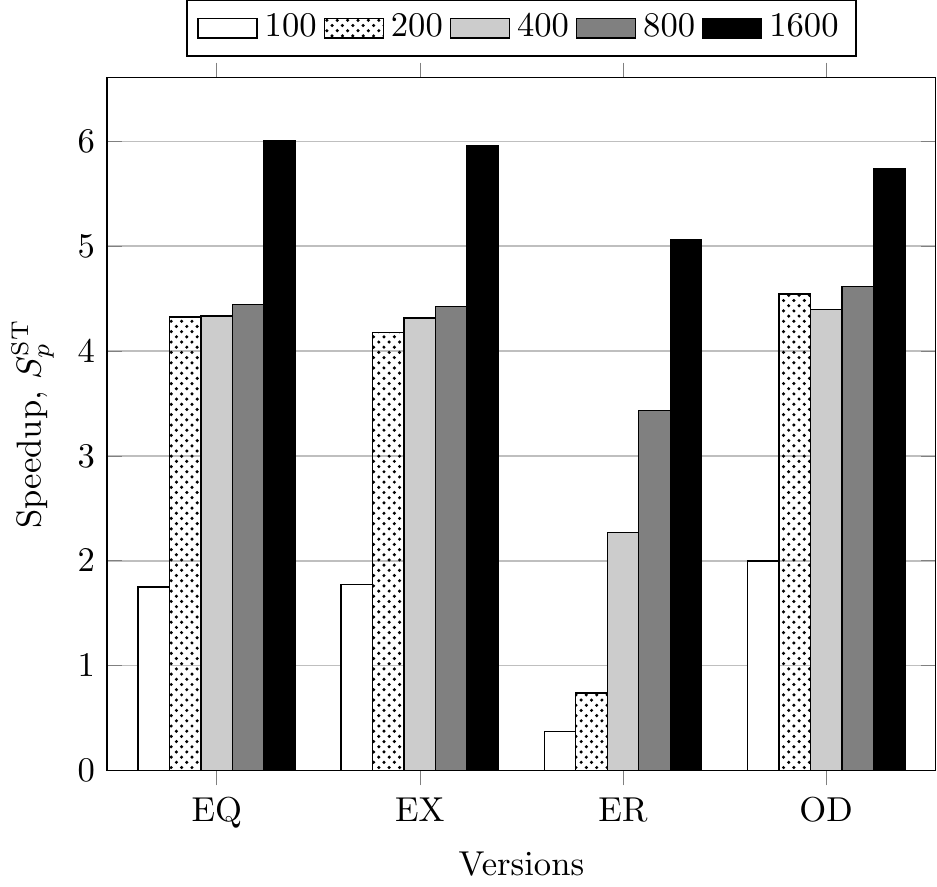}} \\
\subfloat[Speedup against NL, param. set 2.\label{fig:speedups:nlv2}]{\includegraphics[width=0.48\linewidth]{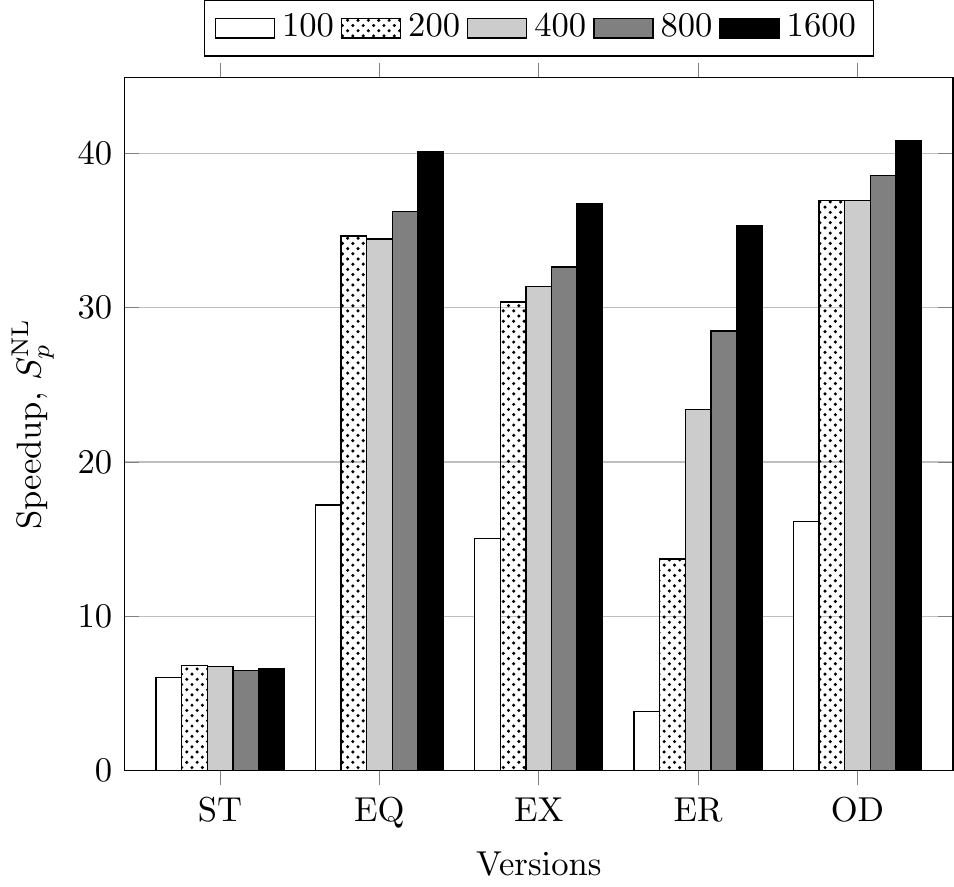}}\quad
\subfloat[Speedup against ST, param. set 2.\label{fig:speedups:jv2}]{\includegraphics[width=0.48\linewidth]{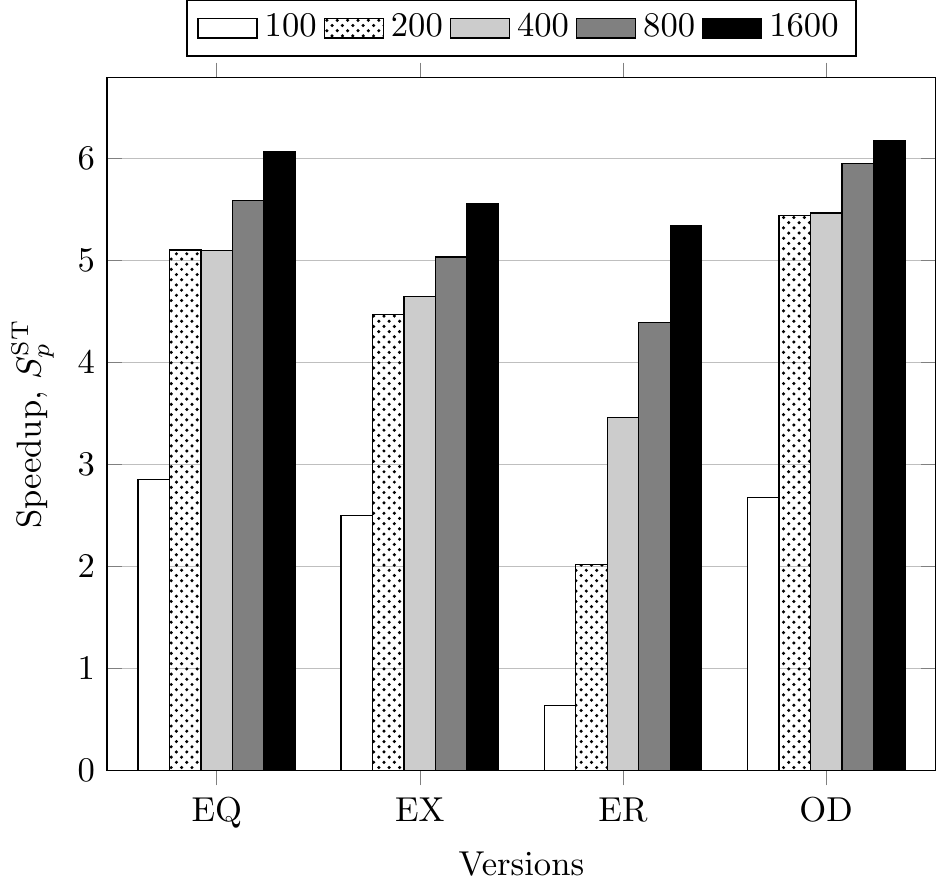}} \\
\caption{Speedup for parallel variants with 12 workers against single\hyp{}thread versions, with $b=500$ for the OD variant.}
\label{fig:speedups}
\end{figure}

Figure \ref{fig:speedups} shows, for both parameter sets, the speedups of the several parallel variants against the NetLogo and Java single\hyp{}thread versions. Speedup is the execution time of the single\hyp{}threaded versions, either $T^{\text{NL}}_1$ for NetLogo, or $T^{\text{ST}}_1$ for Java single\hyp{}thread, over the execution time of a parallel variant using $p$ workers ($T_p$), as defined by

\begin{equation}
\label{eq:speedup}
S_p^{\text{\{NL,ST\}}}=\frac{T_1^{\text{\{NL,ST\}}}}{T_p}
\end{equation}

\begin{figure}[!t]
\centering
\subfloat[Param. set 1.\label{fig:modsize_scal:1}]{\includegraphics[width=0.48\linewidth]{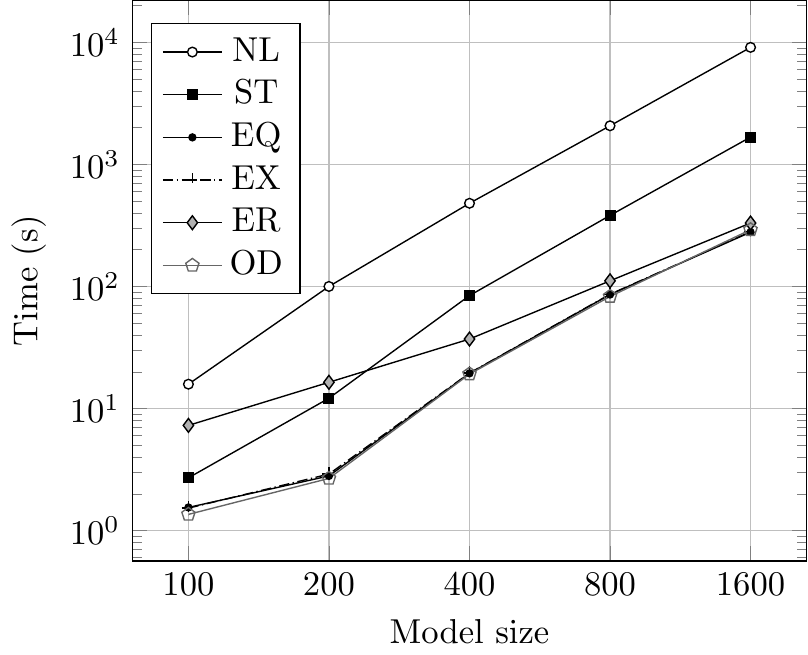}}\quad
\subfloat[Param. set 2.\label{fig:modsize_scal:2}]{\includegraphics[width=0.48\linewidth]{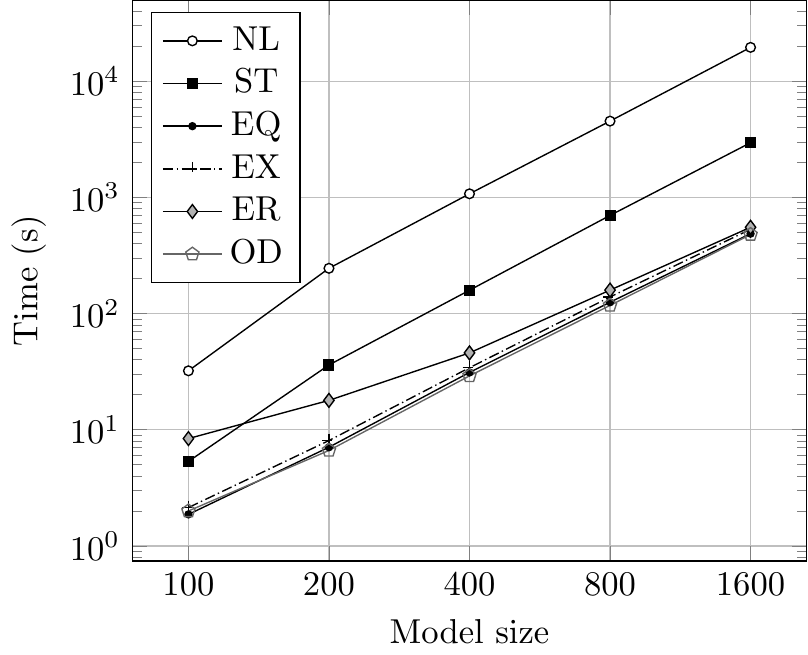}} \\
\caption{Scalability of the different versions for increasing model sizes. Parallel variants are using 12 workers, with $b=500$ for the OD variant.}
\label{fig:modsize_scal}
\end{figure}

Figure \ref{fig:modsize_scal} establishes how the different versions scale with increasingly larger model sizes. The results shown in Figures \ref{fig:speedups} and \ref{fig:modsize_scal} were obtained with 12 worker threads for the Java multithreaded variants, and with $b=500$ for the OD variant. The processor used can concurrently handle 12 threads. With the exception of model size $100$, we find that $N=12$ is the number of workers that offers the best performance. Concerning the OD variant, $b=500$ consistently provides good performance across the different model sizes and parameter sets. Table \ref{tab:times_nj_j} provides more detailed data, including simulation times and relative standard deviations. The latter were consistently small, mainly in the order of 1\%-4\%, and are thus not considered for the remainder of the discussion.

In Figure~\ref{fig:speedups} it is immediately noticeable that the single\hyp{}threaded version of the Java implementation (ST) is $5\times$ to $8\times$ faster than the NetLogo (NL) implementation. Interestingly, the speedup is very consistent for the several tested sizes and for both parameter sets. Although NetLogo is getting considerably more efficient in the last few years \cite{lytinen2012evolution}, implementing an ABM directly in a programming language still seems to provide better performance. Observing the results for the multithreaded Java variants (EQ, EX, ER and OD), speedups up to $40$ further validate that hypothesis. 

The Java implementations with equal work distribution, EQ and EX, offer similar performance, with a slight advantage to the former. The ordered agent insertion performed by the EX variant takes a small toll, but the advantage of repeatability clearly offsets that issue.

It is also obvious that the potential benefits offered by the ER variant, namely the avoidance of cell\hyp{}level synchronization during agent movement, do not outweigh the overhead of row\hyp{}level synchronization. However, as shown in Figures \ref{fig:speedups} and \ref{fig:modsize_scal}, the performance of ER improves for larger model sizes. Nonetheless, the EX variant also offers repeatability, while being faster for the tested model sizes.  The ER variant is actually slower than the ST version for smaller simulations. The only selling point for ER would be if it could outperform other implementations for very large model sizes not tested in this work. This is suggested by Figure~\ref{fig:modsize_scal}, which shows that ER is the only variant that decreases relative simulation time with increasing model sizes.

The OD variant offers the best speedups in most of the test cases. The problem with the OD strategy is that it does not offer repeatable simulations, although a more general solution for this problem is discussed in section \ref{subsubsec:java:rng}. A performance analysis specific for the OD version is presented in section \ref{subsec:odanalysis}.

Generally, as shown in Figure~\ref{fig:speedups}, the speedup of the multithreaded variants improves for larger model sizes. This is clearer for parameter set 2, which yields simulations with more agents. Larger models, in grid size or number of agents, tend to reduce the overhead of parallelization \cite{goldsby2013multithreaded}. The only exception is for model size $200$ with parameter set 1, where all pure Java versions offer substantial speedups versus the NL version (see Figure~\ref{fig:speedups:nlv1}). Figure~\ref{fig:modsize_scal:1} shows this variation can be attributed to consistently faster than expected execution of the Java implementations (except ER) for this size. Although the variation is unexpected, the individual run times do not show much variance between themselves (as shown in Table~\ref{tab:times_nj_j}), so this behavior is consistent and not attributable to outliers. With model size $1600$, most parallel variants offer speedups up to 6 against the ST version, denoting efficient usage of the six\hyp{}core hyper\hyp{}threaded Intel processor. If we consider the number of LPs as the performance target, however, near ideal speedup, $S_p^{\text{ST}} \approx 12$, is harder to achieve because each pair of LPs shares execution resources from a single hyper\hyp{}threaded core. 

\subsection{Varying the number of worker threads}
\label{subsec:results:varthreads}

\begin{figure}[!t]
\centering
\subfloat[Size 100, param. set 1.]{\includegraphics[width=0.48\linewidth]{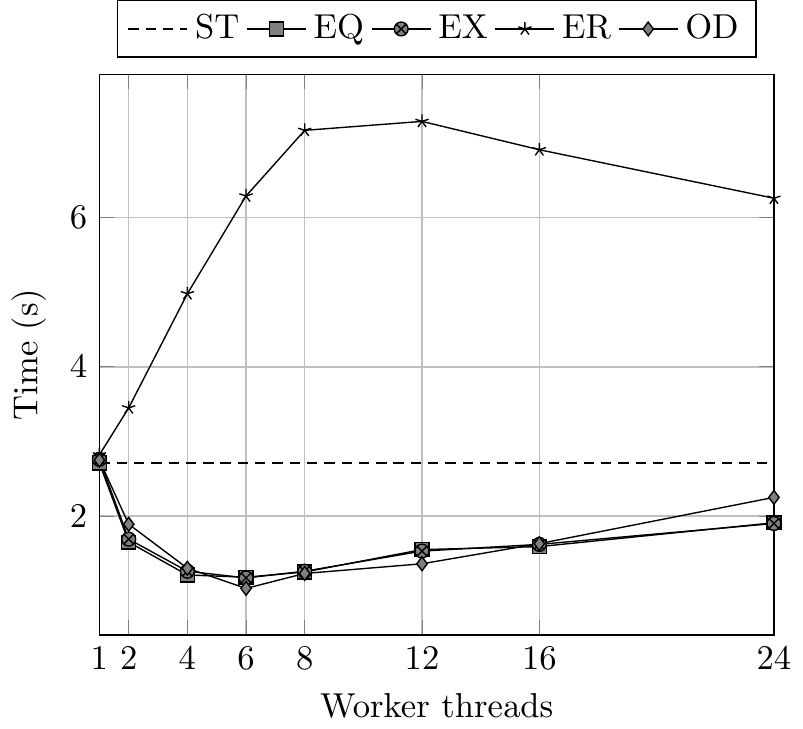}} \,
\subfloat[Size 100, param. set 2.]{\includegraphics[width=0.48\linewidth]{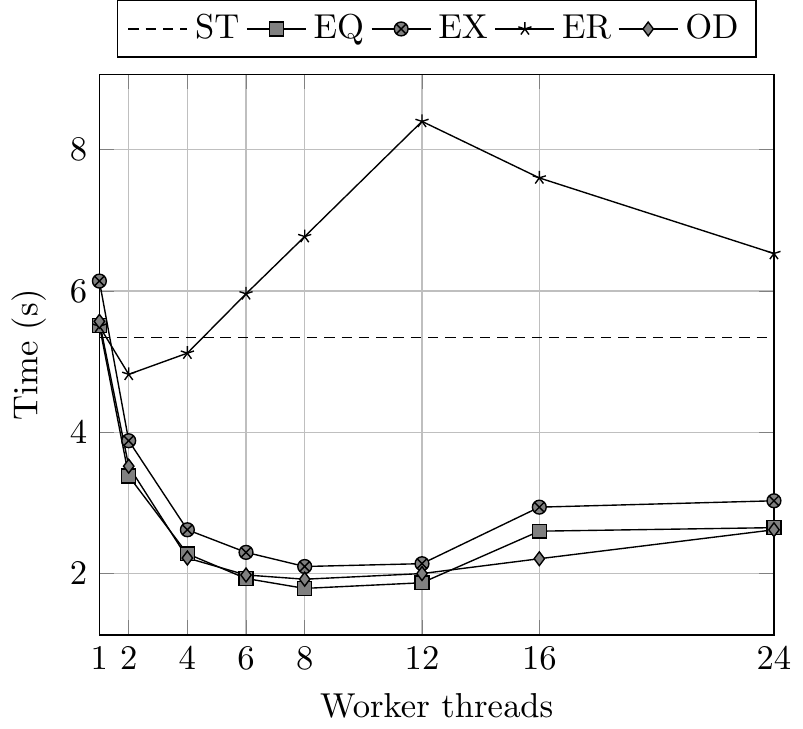}} \,
\subfloat[Size 400, param. set 1.]{\includegraphics[width=0.48\linewidth]{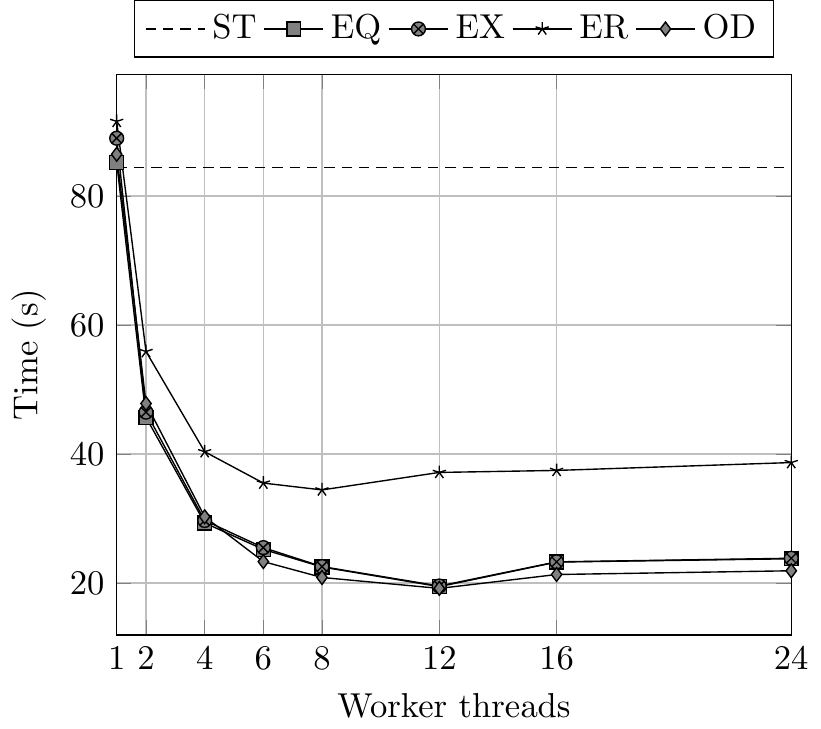}} \,
\subfloat[Size 400, param. set 2.]{\includegraphics[width=0.48\linewidth]{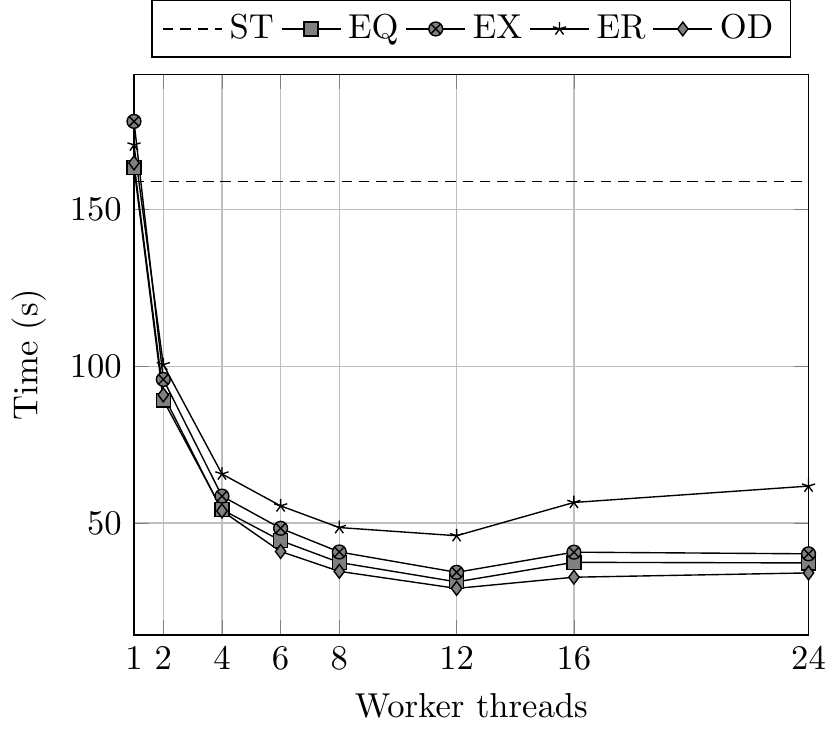}} \,
\subfloat[Size 1600, param. set 1.]{\includegraphics[width=0.48\linewidth]{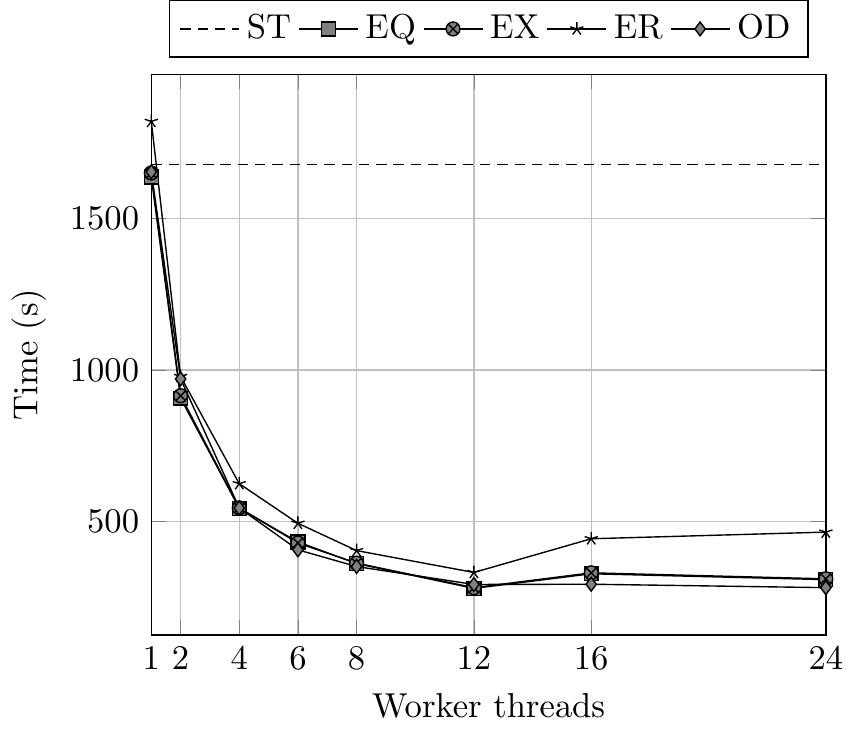}}\,
\subfloat[Size 1600, param. set 2.]{\includegraphics[width=0.48\linewidth]{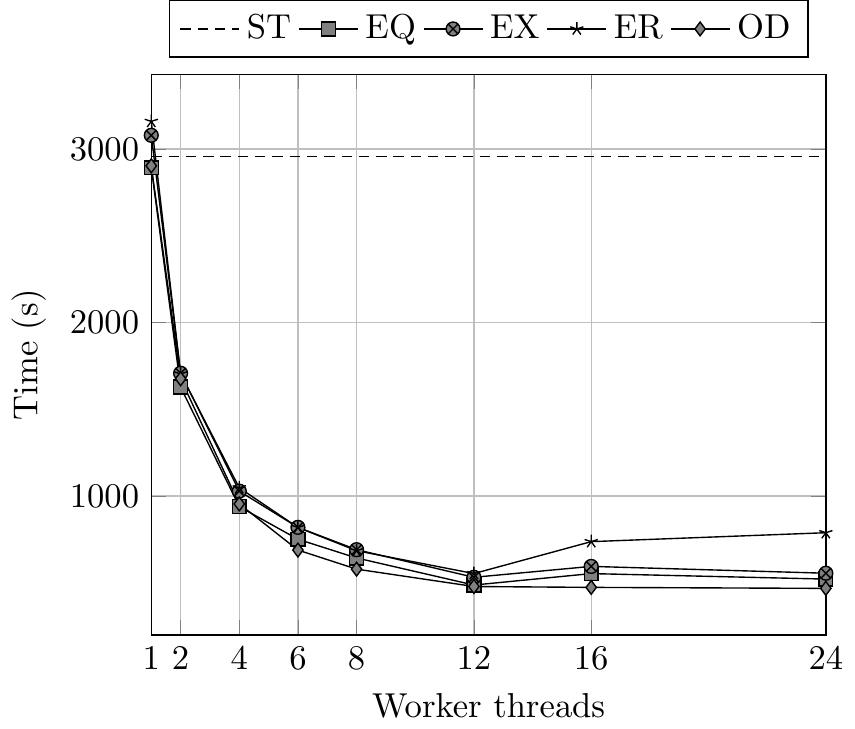}} 
\caption{Simulation time versus number of workers for both parameter sets and model sizes $100$, $400$ and $1600$, with $b=500$ for the OD variant. Time for single\hyp{}thread (ST) implementation shown as reference.}
\label{fig:jthreadvary}
\end{figure}

Figure~\ref{fig:jthreadvary} displays, for the two parameter sets and selected model sizes ($100$, $400$ and $1600$), how the simulation time is affected by varying the number of worker threads in the multithreaded variants. Except for model size $100$, the optimal number of workers is 12, which is in accordance with the number of threads directly supported by the hardware. As a general tendency, larger model sizes benefit more from higher number of workers. The same can be said of parameter set 2, which, having considerably more agents in action, also profits from having more workers. For smaller sizes and/or lower number of agents, each worker thread has less work to do and thus workers spend a larger percentage of time waiting in synchronization points. The OD variant seems to scale better with more workers, most likely because workers process work as it becomes available, spending relatively less time waiting.

\subsection{Analysis of the OD parallelization strategy}
\label{subsec:odanalysis}

\begin{figure}[!t]
\centering
\subfloat[Param. set 1.]{\includegraphics[width=0.48\linewidth]{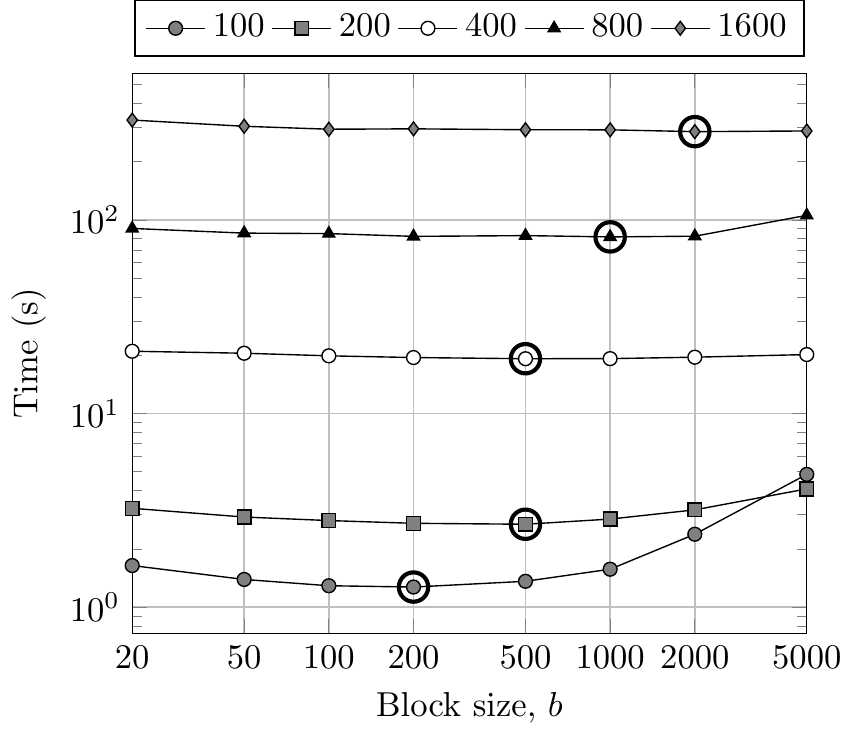}} \,
\subfloat[Param. set 2.]{\includegraphics[width=0.48\linewidth]{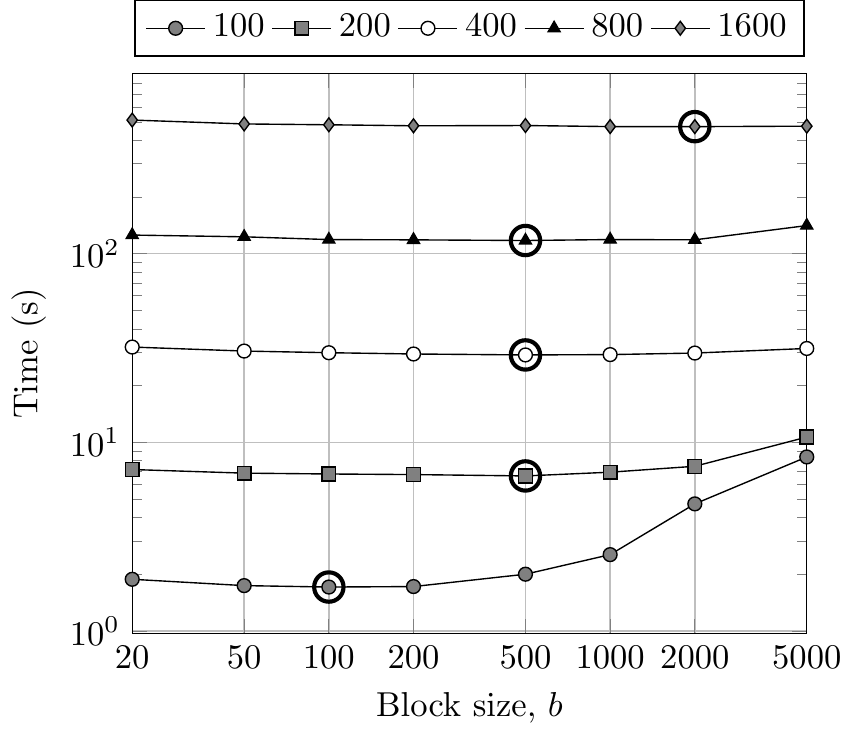}} \,
\caption{OD performance with 12 workers for the two parameter sets and different model sizes. Best performance marked with bold circle.}
\label{fig:jod_perf}
\end{figure}

The OD parallelization strategy deserves a more thorough analysis because of the block size parameter, $b$. Figure~\ref{fig:jod_perf} shows, for the two parameter sets and different model sizes, how simulation times vary with several values of $b$. The number of workers is set to 12. The values of $b$ that offer faster simulation times are marked with a bold circle. Although there is a slight tendency for higher values of $b$ to be associated with larger model sizes, what stands out from Figure~\ref{fig:jod_perf} is that performance is not much affected by $b$, with the exception of size $100$; in this case, larger values of $b$ divide the environment in sizeable sections, such that some workers never get any work to do. For example, a $100 \times 100$ grid has $\num{10000}$ grid cells, and if divided into sections of $\num{5000}$ cells, only two worker threads will have work to do.

Generally, best or close to best simulation times are obtained with $b=500$ across all tested model sizes and parameter sets. As such, this is the value used when comparing OD with the remaining versions.

\subsection{Statistical analysis of model output data}
\label{sec:statanalysis}

The $p$-values obtained from the Kruskal\hyp{}Wallis test by comparing the selected focal measures of the different PPHPC versions for all combinations of parameter sets and model sizes are provided in Table~\ref{tab:j_nl_stats}. In a total of 360 $p$-values (36 focal measures, 2 parameter sets, 5 model sizes), 28 are bellow the $0.05$ significance level; of these, only nine are bellow $0.01$. However, there are two low $p$-values that stand\hyp{}out for $\min \mean{E}^s_i$, associated with model sizes 200 and 800, parameter set 1, suggesting that at least one of the implementations (i.e. samples) produced significantly different minimum values for mean prey energy. Taking a closer look at the distributional behavior of the $\min \mean{E}^s_i$ focal measure for parameter set 1 \cite{fachada2015template}, we note that this value always occurs at iteration zero, the initial state of the simulation (i.e., $\argmin \mean{E}^s_i=0$). Additionally, as described in section \ref{sec:mm:impl}, the EQ and EX parallelization strategies share the same work provider for agent initialization, which means, that for a given PRNG seed, both strategies generate initial agents with the exact same energy. As such, if the initial prey energy is somewhat different from usual for one of these strategies, it will be so for both. Consequently, the Kruskal\hyp{}Wallis test compares two samples with somewhat different observations (EQ and EX) from the remaining four (NL, ST, ER and OD), which is the case for sizes 200 and 800. If we remove EQ and EX from the test, the $p$-values rise to 0.128 and 0.002 for sizes 200 and 800, respectively, in line with the remaining results. Other than this, the $p$-values do not appear to follow any trend or pattern, e.g. smaller $p$-values do not seem to be associated with any particular focal measure, parameter set or model size.

From these results, it is possible to conclude that all realizations appear to produce similar dynamic behavior. Thus, model parallelization does not seem to have introduced any observable bias for the tested parameter sets.

\section{Conclusions}
\label{sec:conclusions}

In this paper, a Java implementation of the PPHPC model, providing several multithreaded parallelization strategies, was proposed. The model captures important characteristics of SABMs, such as agent movement and local agent interactions. Three conclusions are drawn from this study: 

\begin{enumerate}

\item SABM parallelization, if done carefully, can yield considerable performance gains, with speedups up to 40 on a six\hyp{}core hyper\hyp{}threaded processor, and maintain statistical accuracy with the original serial model. While developing models in NetLogo is much simpler than directly using Java or other programming languages, it comes at a considerable cost in terms of performance. 

\item Different parallelization strategies offer specific trade\hyp{}offs in terms of performance and simulation reproducibility. For example, the generally most efficient parallelization strategy, OD, does not allow for reproducible simulations and exposes additional complexity due to the additional block size parameter. The EX variant, however, does support reproducible simulations and is not far off in terms of performance.

\item PPHPC was shown to be a valid reference model for comparing distinct implementations or parallelization strategies, from both performance and statistical accuracy perspectives.

\end{enumerate}

The Java implementation of PPHPC and the respective parallelization strategies can be subject to additional experimentation, e.g. in machines with higher core counts or by exploring additional parameter sets. Nonetheless, future implementations in distinct architectures (e.g. GPU, FPGA, distributed memory) or programming languages, should provide an opportunity to further evaluate how can SABMs be appropriately implemented in each case.

\section{Acknowledgments}

This work was supported by the Fundação para a Ciência e a Tecnologia (FCT) projects UID/EEA/50009/2013, UID/MAT/04561/2013 and (P. RD0389) Incentivo/EEI/LA0009/2014, and partially funded with grant SFRH/BD/48310/2008, also from FCT. The author Vitor V. Lopes acknowledges the financial support from the Prometeo project of SENESCYT (Ecuador).

\bibliographystyle{abbrv}


\begin{table}[t]
\centering
\resizebox{\columnwidth}{!}{%
\begin{tabular}{ccrrrrrrrr}
\toprule

\multirow{2}{*}{Version} & \multirow{2}{*}{Size} & \multicolumn{4}{c}{Param. set 1} & \multicolumn{4}{c}{Param. set 2} \\
\cmidrule(r){3-6} \cmidrule(r){7-10} 
 & & $\mean{t}(\text{s})$ & $s(\%)$ & $S_p^{\text{NL}}$ & $S_p^{\text{ST}}$ & $\mean{t}(\text{s})$ & $s(\%)$ & $S_p^{\text{NL}}$ & $S_p^{\text{ST}}$  \\
\midrule
\multirow{5}{*}{NL}
 & 100 & \num{    15.86} & \num{  2.26}& \num{   1.00} & \num{   0.17} & \num{    32.18} & \num{  2.13}& \num{   1.00} & \num{   0.17}  \\
 & 200 & \num{   100.25} & \num{  1.25}& \num{   1.00} & \num{   0.12} & \num{   245.38} & \num{  0.61}& \num{   1.00} & \num{   0.15}  \\
 & 400 & \num{   481.48} & \num{  1.25}& \num{   1.00} & \num{   0.18} & \num{  1074.21} & \num{  0.34}& \num{   1.00} & \num{   0.15}  \\
 & 800 & \num{  2077.10} & \num{  0.47}& \num{   1.00} & \num{   0.18} & \num{  4536.90} & \num{  0.51}& \num{   1.00} & \num{   0.15}  \\
 & 1600 & \num{  9115.80} & \num{  1.03}& \num{   1.00} & \num{   0.18} & \num{ 19559.30} & \num{  0.46}& \num{   1.00} & \num{   0.15}  \\
\midrule
\multirow{5}{*}{ST}
 & 100 & \num{     2.71} & \num{  0.82}& \num{   5.85} & \num{   1.00} & \num{     5.34} & \num{  0.96}& \num{   6.03} & \num{   1.00}  \\
 & 200 & \num{    12.17} & \num{  1.80}& \num{   8.24} & \num{   1.00} & \num{    36.12} & \num{  0.49}& \num{   6.79} & \num{   1.00}  \\
 & 400 & \num{    84.37} & \num{  3.35}& \num{   5.71} & \num{   1.00} & \num{   158.95} & \num{  0.30}& \num{   6.76} & \num{   1.00}  \\
 & 800 & \num{   382.63} & \num{  1.32}& \num{   5.43} & \num{   1.00} & \num{   699.59} & \num{  0.52}& \num{   6.49} & \num{   1.00}  \\
 & 1600 & \num{  1677.82} & \num{  4.67}& \num{   5.43} & \num{   1.00} & \num{  2957.20} & \num{  4.15}& \num{   6.61} & \num{   1.00}  \\
\midrule
\multirow{5}{*}{EQ}
 & 100 & \num{     1.55} & \num{  1.62}& \num{  10.24} & \num{   1.75} & \num{     1.87} & \num{  1.53}& \num{  17.21} & \num{   2.85}  \\
 & 200 & \num{     2.81} & \num{  4.01}& \num{  35.61} & \num{   4.32} & \num{     7.08} & \num{  1.78}& \num{  34.64} & \num{   5.10}  \\
 & 400 & \num{    19.46} & \num{  1.10}& \num{  24.74} & \num{   4.34} & \num{    31.17} & \num{  0.66}& \num{  34.46} & \num{   5.10}  \\
 & 800 & \num{    86.08} & \num{  4.95}& \num{  24.13} & \num{   4.45} & \num{   125.27} & \num{  3.32}& \num{  36.22} & \num{   5.58}  \\
 & 1600 & \num{   279.23} & \num{  1.45}& \num{  32.65} & \num{   6.01} & \num{   487.34} & \num{  1.74}& \num{  40.14} & \num{   6.07}  \\
\midrule
\multirow{5}{*}{EX}
 & 100 & \num{     1.53} & \num{  1.90}& \num{  10.39} & \num{   1.78} & \num{     2.14} & \num{  2.75}& \num{  15.06} & \num{   2.50}  \\
 & 200 & \num{     2.91} & \num{  3.69}& \num{  34.40} & \num{   4.18} & \num{     8.08} & \num{  1.74}& \num{  30.37} & \num{   4.47}  \\
 & 400 & \num{    19.56} & \num{  1.54}& \num{  24.62} & \num{   4.31} & \num{    34.22} & \num{  1.54}& \num{  31.40} & \num{   4.65}  \\
 & 800 & \num{    86.49} & \num{  6.31}& \num{  24.01} & \num{   4.42} & \num{   138.99} & \num{  4.29}& \num{  32.64} & \num{   5.03}  \\
 & 1600 & \num{   281.57} & \num{  1.95}& \num{  32.37} & \num{   5.96} & \num{   531.96} & \num{  0.99}& \num{  36.77} & \num{   5.56}  \\
\midrule
\multirow{5}{*}{ER}
 & 100 & \num{     7.29} & \num{  4.46}& \num{   2.18} & \num{   0.37} & \num{     8.39} & \num{  1.76}& \num{   3.83} & \num{   0.64}  \\
 & 200 & \num{    16.44} & \num{  4.68}& \num{   6.10} & \num{   0.74} & \num{    17.91} & \num{  1.41}& \num{  13.70} & \num{   2.02}  \\
 & 400 & \num{    37.16} & \num{  0.55}& \num{  12.96} & \num{   2.27} & \num{    45.91} & \num{  0.62}& \num{  23.40} & \num{   3.46}  \\
 & 800 & \num{   111.45} & \num{  3.02}& \num{  18.64} & \num{   3.43} & \num{   159.25} & \num{  2.02}& \num{  28.49} & \num{   4.39}  \\
 & 1600 & \num{   331.77} & \num{  1.06}& \num{  27.48} & \num{   5.06} & \num{   553.44} & \num{  1.45}& \num{  35.34} & \num{   5.34}  \\
\midrule
\multirow{5}{*}{OD}
 & 100 & \num{     1.36} & \num{  1.16}& \num{  11.70} & \num{   2.00} & \num{     2.00} & \num{  1.66}& \num{  16.13} & \num{   2.68}  \\
 & 200 & \num{     2.68} & \num{  2.61}& \num{  37.42} & \num{   4.54} & \num{     6.64} & \num{  1.64}& \num{  36.95} & \num{   5.44}  \\
 & 400 & \num{    19.19} & \num{  1.04}& \num{  25.09} & \num{   4.40} & \num{    29.09} & \num{  0.42}& \num{  36.93} & \num{   5.46}  \\
 & 800 & \num{    82.94} & \num{  2.73}& \num{  25.04} & \num{   4.61} & \num{   117.62} & \num{  2.55}& \num{  38.57} & \num{   5.95}  \\
 & 1600 & \num{   292.16} & \num{  2.91}& \num{  31.20} & \num{   5.74} & \num{   478.83} & \num{  1.95}& \num{  40.85} & \num{   6.18}  \\
\bottomrule
\end{tabular}
}
\caption{\label{tab:times_nj_j}Times and speedups for the different versions using both
 parameter sets and tested model sizes. The $\mean{t}(\text{s})$ column specifies the mean simulation time for 
 each version and model size combination. The $s(\%)$ column shows the associated relative standard deviation, 
 given by $100 \cdot s/\mean{t}(\text{s})$, where $s$ is the sample standard deviation. The $S_p^{\text{NL}}$ and 
 $S_p^{\text{ST}}$  columns display the speedups verified against the NetLogo and Java single\hyp{}thread versions, 
 respectively. The number of workers, $p$, is set to 12 for the parallel variants, and 1 for the NL and ST 
 versions. For the OD variant, $b=500$.}
\end{table}

\begin{table}[t]
\centering
\resizebox{\columnwidth}{!}{%
\begin{tabular}{clrrrrrrrrrr}
\toprule
\multirow{2}{*}{Output} & \multirow{2}{*}{Stat.} & \multicolumn{5}{c}{Param. set 1} & \multicolumn{5}{c}{Param. set 2} \\
\cmidrule(r){3-7} \cmidrule(r){8-12}
 & & 100 & 200 & 400 & 800 & 1600 & 100 & 200 & 400 & 800 & 1600 \\

\midrule
\multirow{6}{*}{$P_i^s$}
 & $\max$ &          0.777 &  \uline{0.046} &          0.283 &          0.833 &          0.832 &  \uline{0.018} &          0.203 &          0.894 &          0.584 &          0.418 \\ 
 & $\argmax$ &          0.785 &          0.456 &          0.990 &  \uline{0.030} &          0.575 &          0.685 &          0.600 &          0.785 &          0.445 &          1.000 \\ 
 & $\min$ &          0.776 &  \uline{0.029} &          0.086 & \uuline{0.007} &          0.161 &          0.191 &          0.323 &          0.539 &          1.000 &          1.000 \\ 
 & $\argmin$ &          0.546 &          0.679 &          0.542 &          0.085 &          0.348 &          0.555 &          0.169 &          0.539 &          1.000 &          1.000 \\ 
 & $\mean{X}^{\text{ss}}$ &          0.489 &          0.984 &          0.298 &          0.533 &          0.976 &          0.683 &          0.079 &          0.204 &          0.886 &          0.199 \\ 
 & $S^{\text{ss}}$ &          0.054 &          0.610 &          0.572 &          0.790 &  \uline{0.025} &          0.729 &          0.194 &          0.530 &          0.387 &          0.159 \\ 
\midrule
\multirow{6}{*}{$P_i^w$}
 & $\max$ &          0.062 &          0.981 &          0.449 &          0.484 &          0.685 &          0.436 &          0.533 &          0.205 &          0.249 & \uuline{0.008} \\ 
 & $\argmax$ &          0.944 &          0.494 &          0.658 &          0.579 &          0.469 &          0.443 &          0.667 &          0.243 &          0.793 &          0.698 \\ 
 & $\min$ &          0.615 &          0.168 &          0.447 &          0.675 &          0.438 &          0.081 &          0.720 &          0.483 &          0.111 &          0.483 \\ 
 & $\argmin$ &          0.381 &          0.407 &          0.382 &          0.410 &          0.630 &          0.846 &          0.444 &          0.084 &          0.362 &          0.707 \\ 
 & $\mean{X}^{\text{ss}}$ &          0.428 &          0.995 &          0.815 &          0.729 &          0.499 &          0.785 &          0.177 &          0.893 &          0.632 &          0.636 \\ 
 & $S^{\text{ss}}$ &          0.161 &          0.795 &          0.840 &          0.604 & \uuline{0.003} &          0.700 &          0.268 &          0.667 &          0.350 &          0.185 \\ 
\midrule
\multirow{6}{*}{$P_i^c$}
 & $\max$ &          0.717 &          0.565 &          0.144 &          0.054 &  \uline{0.025} &          0.298 &          0.542 &          0.129 &          0.583 &  \uline{0.036} \\ 
 & $\argmax$ &          0.383 &          0.749 &          0.582 &          0.416 &          1.000 &  \uline{0.050} &          0.534 &          0.235 &          0.739 &          0.463 \\ 
 & $\min$ &          0.766 &          0.122 &          0.593 &          0.379 &          0.854 &          0.118 &          0.209 &          0.644 &          0.426 &          0.217 \\ 
 & $\argmin$ &          0.942 &          0.802 &          0.067 &          0.896 &          0.823 &          0.822 &          0.862 &          0.413 &          0.302 &          0.671 \\ 
 & $\mean{X}^{\text{ss}}$ &          0.566 &          0.970 &          0.386 &          0.488 &          0.887 &          0.956 &          0.453 &          0.067 &          0.421 &          0.367 \\ 
 & $S^{\text{ss}}$ &          0.073 &          0.543 &          0.644 &          0.785 &  \uline{0.011} &          0.723 &          0.255 &          0.644 &          0.358 &          0.123 \\ 
\midrule
\multirow{6}{*}{$\mean{E}^s_i$}
 & $\max$ &          0.595 &          0.183 &          0.410 &          0.797 &          0.718 &          0.489 &          0.111 &          0.538 & \uuline{0.004} &          0.115 \\ 
 & $\argmax$ &          0.970 &          0.169 &          0.443 &          0.180 &          0.369 &          0.940 &          0.520 &          0.819 &          0.348 &          1.000 \\ 
 & $\min$ & \uuline{0.003} & \uuline{\num[output-exponent-marker=\text{e}]{2e-05}} &          0.142 & \uuline{\num[output-exponent-marker=\text{e}]{4e-06}} &          0.099 &          0.400 &          0.077 &          0.967 &          0.102 &  \uline{0.012} \\ 
 & $\argmin$ &          1.000 &          1.000 &          1.000 &          1.000 &          1.000 &          0.705 &          0.485 &          0.871 &          0.566 &          0.627 \\ 
 & $\mean{X}^{\text{ss}}$ &          0.901 &          0.236 &          0.909 &          0.981 &          0.987 &          0.451 &  \uline{0.024} &          0.276 &          0.372 &          0.216 \\ 
 & $S^{\text{ss}}$ &          0.242 &          0.633 &          0.200 &          0.480 &          0.250 &          0.642 &          0.458 &          0.395 &          0.425 &          0.250 \\ 
\midrule
\multirow{6}{*}{$\mean{E}^w_i$}
 & $\max$ &          0.508 &          0.664 &          0.696 &          0.672 &          0.716 &          0.269 &          0.101 &          0.619 &          0.944 &          0.331 \\ 
 & $\argmax$ &          0.338 &          0.687 &          0.156 &          0.247 &          0.467 &          0.499 &          0.620 &          0.057 &          0.974 &          0.371 \\ 
 & $\min$ &          0.518 &  \uline{0.019} & \uuline{0.005} &          0.143 &  \uline{0.020} &          0.074 &          0.973 &          0.079 &          0.145 &          0.591 \\ 
 & $\argmin$ &          0.109 &          0.963 &          0.514 &          0.425 &          0.278 &          0.598 &  \uline{0.033} &          0.459 &          0.317 &          0.949 \\ 
 & $\mean{X}^{\text{ss}}$ &  \uline{0.026} &          0.320 &          0.224 &          0.204 &          0.686 &          0.971 & \uuline{0.010} &          0.345 &          0.876 &          0.212 \\ 
 & $S^{\text{ss}}$ &          0.547 &          0.897 &          0.741 &          0.986 &          0.061 &          0.663 &          0.400 &          0.665 &          0.457 &          0.115 \\ 
\midrule
\multirow{6}{*}{$\mean{C}_i$}
 & $\max$ &          0.795 &          0.129 &          0.622 &          0.313 &          0.742 &  \uline{0.045} &          0.823 &          0.270 &          0.717 &          0.118 \\ 
 & $\argmax$ &          0.674 &          0.792 &  \uline{0.044} &          0.996 &          0.772 &          0.257 &          0.581 &          0.234 &          0.735 &          0.276 \\ 
 & $\min$ &          0.612 &          0.712 &          0.164 &          0.118 &          0.056 &          0.261 &          0.586 &          0.165 &          0.640 &  \uline{0.037} \\ 
 & $\argmin$ &          0.709 &          0.834 &          1.000 &          1.000 &          1.000 &          0.078 &          0.570 &          0.547 &          0.901 &          0.904 \\ 
 & $\mean{X}^{\text{ss}}$ &          0.585 &          0.975 &          0.365 &          0.509 &          0.880 &          0.966 &          0.423 &          0.072 &          0.404 &          0.326 \\ 
 & $S^{\text{ss}}$ &          0.073 &          0.539 &          0.644 &          0.776 &  \uline{0.010} &          0.702 &          0.250 &          0.647 &          0.367 &          0.126 \\ 
\bottomrule

\end{tabular}}
\caption{\label{tab:j_nl_stats}$P$-values for the Kruskal\hyp{}Wallis statistical test which tests the null hypothesis that the specified focal measures from the six PPHPC versions are drawn from the same distribution. Smaller $p$-values (lower than $0.05$ or $0.01$) question the null hypothesis, indicating that at least one implementation produced an output significantly different from the others. Values lower than $0.05$ are underlined, while values lower than $0.01$ are double\hyp{}underlined. The number of workers, $p$, is set to 12 for the parallel implementations. For the OD implementation, $b=500$.}
\end{table}

\end{document}